Entire latex paper, to appear in Russian journal Physics of Elementary 
Particles and Atomic Nuclei
\documentstyle[12pt]{report}
\def\lg{{\mathchoice{~\raise.58ex\hbox{$<$}\mkern-14.8mu\lower.52ex\hbox{$>$}~}
	            {~\raise.58ex\hbox{$<$}\mkern-14.8mu\lower.52ex\hbox{$>$}~}
		    {\raise.59ex\hbox{{$\scriptscriptstyle <$}}\mkern-12.8mu%
		     \lower.01ex\hbox{{$\scriptscriptstyle >$}}}   {} 	}} 
\def\gl{{\mathchoice{~\raise.58ex\hbox{$>$}\mkern-12.8mu\lower.52ex\hbox{$<$}~}
                    {~\raise.58ex\hbox{$>$}\mkern-12.8mu\lower.52ex\hbox{$<$}~}
		    {\raise.62ex\hbox{{$\scriptscriptstyle >$}}\mkern-12.0mu%
		     \lower.05ex\hbox{{$\scriptscriptstyle <$}}}  {} 	}}   

\begin{document}
\vspace{3cm}

\begin{center}

{\Large \bf Topics in the Transport Theory \\ [5mm]
of Quark-Gluon Plasma} \\

\vspace{3cm}

Stanis\l aw Mr\' owczy\' nski\footnote{E-mail: MROW@FUW.EDU.PL} \\[3mm]
\it
So\l tan Institute for Nuclear Studies,\\
ul. Ho\. za 69, PL - 00-681 Warsaw, Poland \\
and Institute of Physics, Pedagogical University,\\
ul. Konopnickiej 15, PL - 25-406 Kielce, Poland\\
\vspace{2cm}
\rm
\begin{minipage}{13cm}
\baselineskip=12pt

{\small \qquad A few topics of the transport theory of quark-gluon plasma
are reviewed. A derivation of the transport equations form the underlaying
dynamical theory is discussed within the $\phi^4$ model. Peculiarities
of the kinetic equations of quarks and gluons are considered and the
plasma (linear) response to the color field is studied. The chromoelectric 
tensor permeability is found and the plasma oscillations are discussed. 
Finally, the filamentation instability in the strongly anisotropic parton
system from ultrarelativistic heavy-ion collisions is discussed in detail.}

\end{minipage}
\end{center}

\tableofcontents

\chapter{Introduction}

The quark-gluon plasma (QGP) is a macroscopic system of deconfined quarks 
and gluons. The very existence of QGP at a sufficiently large temperature
and/or baryon density is basically an unavoidable consequence of the 
quantum chromodynamics (QCD) which is a dynamical theory of strong 
interactions (see e.g. \cite{Ynd83}). The plasma has been present 
in the early Universe and presumably can be found in the compact stellar 
objects. Of particular interest however is the generation of QGP in 
relativistic heavy-ion collisions which has been actively studied 
theoretically and experimentally \cite{QM96} for over ten years. The 
life time of the plasma produced, if indeed produced, in these 
collisions is not much longer than the characteristic time scale of parton 
processes\footnote{The word {\it parton} is used as a common name of 
quarks and gluons.}. Therefore, QGP can achieve, in the best case, only 
a quasi-equilibrium state and studies of the nonequilibrium phenomena 
are crucial to discriminate the characteristic features of QGP.

The transport or kinetic theory provides a natural framework to study systems
out of thermodynamical equilibrium. Although the theory was initiated 
more than a century ago - Boltzmann derived his famous equation in 1872, 
the theory is still under vital development. Application of the Boltzmann's 
ideas to the systems which are relativistic and of quantum nature is faced with 
difficulties which have been overcome only partially till now. For a review 
see the monography \cite{Gro80}. In the case of the quark-gluon plasma 
specific difficulties appear due to the system non-Abelian dynamics. 
Nevertheless, the transport theory approach to QGP is in fast progress and 
some interesting results have been already found. 
 
The aim of this article is to review a few topics of the QGP transport 
theory. The first one is how to derive the transport equations of quarks
and gluons. Since QCD is the underlying dynamical theory, these equations 
should be deduced from QCD. However, the kinetic theory of quarks and gluons 
have been successfully derived from QCD only in the mean-field or 
collisionless limit \cite{Elz89,Bla94}. The derivation of the collision 
terms is still an open question. We discuss here the issue within the 
dynamical model which is much simpler than QCD. Namely, we consider the self 
interacting scalar fields with the quartic interaction term. Then, one can 
elucidate the essence of the derivation problem.

In the third chapter we present the transport equations of quarks and
gluons obtained in the mean-field limit. The equations are supplemented
by the collision terms which are justified on the phenomenological ground. 
We briefly discuss the peculiarities of the transport theory of 
quarks and gluons and then consider the locally colorless plasma\footnote{We 
call the plasma {\it locally colorless} if the color four-current vanishes 
at each space-time point. It differs from the terminology used in the 
electron-ion plasma physics, where the plasma is called {\it locally 
neutral} if the electric charge (zero component of electromagnetic 
four-current) is everywhere zero.}. The dynamical content of QCD enters 
here only through the cross-sections of parton-parton interactions. 

The characteristic features of QGP appear when the plasma is not locally 
colorless and consequently it interacts with the chromodynamic mean field. 
The plasma response to such a field is discussed in the fourth chapter,
where the color conductivity and chromoelectric permeability tensors
are found. We also analyze there the oscillations around the global 
thermodynamical equilibrium. 

The parton momentum distribution is expected to be strongly anisotropic 
at the early stage of ultrarelativistic heavy-ion collisions. Then, the
parton system can be unstable with respect to the specific plasma modes.
In the fifth chapter we discuss in detail the mode, which splits the parton 
system into the color current filaments parallel to the beam direction.
We show why the fluctuation which initiates the filamentation can be 
very large and explain the physical mechanism responsible for the fluctuation 
growth. Then, the exponentially growing mode is found as a solution 
of the respective dispersion equation. The characteristic time of the 
instability development is estimated and finally, the possibility to observe 
the color filamentation in nucleus-nucleus collisions at RHIC and LHC 
is considered.

Presenting the QGP transport theory we try to avoid model dependent 
concepts but a very crucial assumption is adopted that the plasma is 
{\it perturbative} i.e. the partons weakly interact with each other. 
As known, QGP becomes perturbative only at the temperatures much greater 
than the QCD scale parameter $\Lambda \cong 200$ MeV, see e.g. \cite{Kal84}.
However, one believes that many results obtained in the framework of 
the perturbative QCD can be extrapolated to the {\it nonperturbative} 
regime.

In the whole article we use the units where $c=k=\hbar =1$. The metric 
tensor is diagonal with $g_{00}=-g_{11}=-g_{22}=-g_{33} = 1$.

\chapter{Derivation of the transport equation in $\phi^4$ model}

The transport equations can be usually derived by means of simple heuristic 
arguments similar to those which were used by Boltzmann when he formulated the
kinetic theory of gases. However, such arguments are insufficient when one 
studies a system of complicated dynamics as the quark-gluon plasma governed 
by QCD. Then, one has to refer to a formal scheme which allows to derive 
the transport equation directly from the underlying quantum field theory. 
The formal scheme is also needed to specify the limits of the kinetic 
approach. Indeed, the derivation shows the assumptions and approximations 
which lead to the transport theory, and hence the domain of its applicability 
can be established.

Until now the transport equations of the QCD plasma have been successfully 
derived in the mean-field limit \cite{Elz89,Bla94} and the structure of 
these equations is well understood \cite{Elz89,Bla94,Mro89,Kel94}. In 
particular, it has been shown that in the quasiequilibrium these equations 
provide \cite{Bla94,Kel94} the so-called hard thermal loops \cite{Bra90}. 
The collisionless transport equations can be applied to the variety of 
problems. However, one needs the collision terms to discuss dissipative 
phenomena. In spite of some efforts \cite{Sel91,Sel94,Gei96}, the general 
form of these terms in the transport equations of the quark-gluon plasma 
remains unknown. 

The so-called Schwinger-Keldysh \cite{Sch61,Kel64} formulation of the quantum field 
theory provides a very promising basis to derive the transport equation
beyond the mean-field limit. Kadanoff and Baym \cite{Kad62} developed the 
technique for nonrelativistic quantum systems which has been further 
generalized to relativistic ones 
\cite{Bez72,Li83,Dan84,Cal88,Mro90,Mro94a,Hen95,Boy96,Gei96,Kle97,Reh98}. 
We mention here only the papers which provide a more or less systematic 
analysis of the collision terms. 

The derivation of the complete QCD transport equations appears to be 
a very difficult task. In particular, the treatment of the massless fields
such as gluons is troublesome. Except the well known infrared divergences 
which plague the perturbative expansion, there is a specific problem 
of nonequilibrium massless fields. The inhomogeneities in the system cause 
the off-mass-shell propagation of particles and then the perturbative 
analysis of the collision terms appears hardly tractable. More specifically, 
it appears very difficult, if possible at all, to express the field 
self-energy as the transition matrix element squared and consequently we loose 
the probabilistic character of the kinetic theory. The problem is absent 
for the massive fields when the system is assumed to be homogeneous at the 
inverse mass or Compton scale. This is a natural assumption within 
the transport theory which anyway deals with the quantities averaged over 
a certain scale which can be identified with the Compton one. 
We have developed \cite{Mro97a} a systematic approach to the transport 
of massless fields, which allows one to treat these fields in a very similar 
manner as the massive ones. The basic idea is rather obvious. The fields 
which are massless in vacuum gain an effective mass in a medium due to 
the interaction. Therefore, the minimal scale at which the transport 
theory works is not an inverse bare mass, which is infinite for massless 
fields, but the inverse effective one. The staring point of the 
perturbative computation should be no longer free fields but the 
interacting ones. In physical terms, we have postulated existence of 
the massive quasiparticles and look for their transport equation.
We have successfully applied the method to the massless scalar fields
\cite{Mro97a}, but the generalization to QCD is far not straightforward 
due to the much reacher quasiparticle spectrum.

To demonstrate the characteristic features of the transport theory 
derivation we discuss in this chapter the simplest nontrivial model
i.e. the real massive fields with the lagrangian density of the form
\begin{equation}\label{lagran}
{\cal L}(x) = {1 \over 2}\partial^{\mu}\phi (x)\partial_{\mu}\phi (x)
- {1 \over 2}m^2 \phi^2(x) - {g \over 4!}\phi^4(x) \;.
\end{equation}
The main steps of the derivation are the following. One defines the contour 
Green function with the time arguments on the contour in a complex time 
plane. This function, which is a key element of the Schwinger-Keldysh 
approach, satisfies the Dyson-Schwinger equation. Assuming the macroscopic 
quasi-homogeneity of the system, one performs the gradient expansion and 
the Wigner transformation. Then, the pair of Dyson-Schwinger equations is  
converted into the transport and mass-shell equations both satisfied by 
the Wigner function. One further computes perturbatively the self energy
which provides the Vlasov and the collisional terms of the transport 
equation. Finally, one defines the distribution function of standard 
probabilistic interpretation and find the transport equation satisfied 
by this function.

\section{Green functions}

The contour Green function is defined as 
$$
i\Delta (x,y) \buildrel \rm def \over 
= \langle  \tilde T \phi (x) \phi (y) \rangle \;, 
$$
where $\langle ....\rangle $ denotes the ensemble average at time $t_0$ 
(usually identified with $-\infty $); $\tilde T$ is the time ordering 
operation along the directed contour shown in Fig.~1. The parameter 
$t_{max}$ is shifted to $+ \infty$ in the calculations. The time arguments 
are complex with an infinitesimal positive or negative imaginary part, 
which locates them on the upper or on the lower branch of the contour. 
The ordering operation is defined as
$$
\tilde T \phi (x) \phi (y) \buildrel \rm def \over = 
\Theta (x_0,y_0)\phi (x) \phi (y) +
\Theta (y_0,x_0)\phi (y) \phi (x) \;,
$$
where $\Theta (x_0,y_0)$ equals 1 if $x_0$ succeeds $y_0$ on the 
contour and  equals 0 when $x_0$  precedes $y_0$. 

If the field is expected to develop a finite expectation value,
as it happens when the symmetry is spontaneously broken, the
contribution $\langle \phi (x)\rangle \langle \phi (y) \rangle$  
is subtracted from the right-hand-side of the equation defining the Green 
function, see e.g. \cite{Mro90,Mro94a}. Then, one concentrates on the 
field fluctuations around the expectation values. Since 
$\langle \phi (x)\rangle$ is expected to vanish in the models
defined by the lagrangians (\ref{lagran}) we neglect this contribution
in the Green function definition. 

We also use four other Green functions with real time arguments:
$$
i\Delta^> (x,y) \buildrel \rm def \over 
= \langle  \phi (x) \phi (y) \rangle \;, 
$$
$$
i\Delta^<  (x,y) \buildrel \rm def \over = 
\langle  \phi (y) \phi (x) \rangle \;, 
$$
$$
i\Delta^c (x,y) \buildrel \rm def \over = 
\langle  T^c \phi (x) \phi (y) \rangle \;, 
$$
$$
i\Delta^a (x,y) \buildrel \rm def \over 
= \langle  T^a \phi (x) \phi (y) \rangle \;, 
$$
where $T^c (T^a)$ prescribes (anti-)chronological time ordering:
$$
T^c \phi (x) \phi (y) \buildrel \rm def \over = 
\Theta (x_0-y_0) \phi (x) \phi (y) +
\Theta (y_0-x_0) \phi (y) \phi c(x) \;,
$$
$$
T^a \phi (x) \phi (y) \buildrel \rm def \over = 
\Theta (y_0-x_0) \phi (x) \phi (y) +
\Theta (x_0-y_0) \phi (y) \phi (x) \;.
$$
These functions are related to the contour Green functions 
in the following manner:
$$
\Delta^c(x,y) \equiv \Delta (x,y) \;\; {\rm for} \;\;
x_0 , \; y_0 \;\; {\rm from \;\; the \;\; upper \;\; branch,} 
$$
$$
\Delta^a(x,y) \equiv \Delta (x,y) \;\; {\rm for} \;\;
x_0 , \; y_0 \;\; {\rm from \;\; the \;\; lower \;\; branch,} 
$$
\begin{eqnarray*}
\Delta^>(x,y) \equiv \Delta (x,y) \;\; {\rm for} \;\;
&x_0&  \; {\rm from \;\; the \;\; upper \;\; branch\;\; and \;\;} \\
&y_0& \;\; {\rm from \;\; the \;\; lower \;\; one,} 
\end{eqnarray*}
\begin{eqnarray*}
\Delta^<(x,y) \equiv \Delta (x,y) \;\; {\rm for} \;\;
&x_0& \;\; {\rm from \;\; the \;\; lower \;\; branch  \;\; and \;\;} \\
&y_0& \;\; {\rm from \;\; the \;\; upper \;\; one. } 
\end{eqnarray*}

It appears convenient to introduce the retarded $(+)$ and advanced
$(-)$ Green functions 
\begin{equation}\label{ret}
\Delta^{\pm} (x,y) \buildrel \rm def \over = \pm \Bigr(
\Delta^>(x,y) - \Delta^<(x,y) \Bigl) \Theta (\pm x_0 \mp y_0) \;.
\end{equation}
One easily finds several identities which directly follow from the 
definitions and relate the Green functions to each other.

$\Delta^c(x,y)$ describes the propagation of disturbance in which 
a single particle is added to the many-particle system in space-time 
point $y$ and then is removed from it in a space-time point $x$. 
An antiparticle disturbance is propagated backward in time. The meaning of 
$\Delta^a(x,y)$ is analogous but particles are propagated backward
in time and antiparticles forward. In the zero density limit 
$\Delta^c(x,y)$ coincides with the Feynman propagator.

The physical meaning of functions $\Delta^>(x,y)$ and 
$\Delta^<(x,y)$ is more transparent when one considers the Wigner 
transform defined as
\begin{equation}\label{Wigner}
\Delta^{\lg}(X,p) \buildrel \rm def \over = \int d^4u e^{ipu}
\Delta^{\lg}(X+{1 \over 2}u,X-{1 \over 2}u) \;.
\end{equation}
Then, the free-field energy-momentum tensor averaged over ensemble can 
be expressed as
\begin{eqnarray*}
t^{\mu \nu}_0(X) \buildrel \rm def \over =
-{1 \over 4} \langle \phi (x)\buildrel \leftrightarrow
\over \partial^{\mu} \buildrel \leftrightarrow
\over \partial^{\nu} \phi (x) \rangle =
\int {d^4p \over (2\pi)^4} p^{\mu} p^{\nu} i\Delta^<(X,p) \;.
\end{eqnarray*}
One recognizes the standard form of the energy-momentum tensor 
in the kinetic theory with the function $i\Delta^<(X,p)$ giving 
the density of particles with four-momentum $p$ in a space-time 
point $X$. Therefore, $i\Delta^<(X,p)$ can be treated as a quantum 
analog of the classical distribution function. Indeed, the function 
$i\Delta^<(X,p)$ is hermitian. However it is not positively definite 
and the probabilistic interpretation is only approximately valid. 
One should also observe that, in contrast to the classical distribution
functions, $i\Delta^<(X,p)$ can be nonzero for the off-mass-shell 
four-momenta. 

\section{Equations of Motion}

The Dyson-Schwinger equations satisfied by the contour Green 
function are
\begin{equation}\label{DS1}
\big[\partial_x^2 + m^2\big] \Delta (x,y) = 
-\delta^{(4)}(x,y) + \int_C d^4x' \Pi(x,x')
\Delta(x',y) \;, 
\end{equation}
\begin{equation}\label{DS2}
\big[\partial_y^2 + m^2\big] \Delta (x,y) = 
-\delta^{(4)}(x,y) + \int_C d^4x' \Delta(x,x')
\Pi(x',y) \;,
\end{equation}
where $\Pi(x,y)$ is the self-energy; the  integration  over  
$x'_0$ is  performed  on  the contour and the function 
$\delta^{(4)}(x,y)$ is defined on the contour as 
\begin{displaymath}
\delta^{(4)}(x,y) = \left\{ \begin{array}{ccl} 
\delta^{(4)}(x-y) \;\;\; & {\rm for} &\;\; x_0 \;, \; y_0 \;\; 
{\rm from \;\; the \;\; upper \;\; branch,} \\
0 \;\;\;\;\;\;\; & {\rm for} &\;\; x_0 \;, \; y_0 \;\; 
{\rm from \;\; the \;\; different \;\; branches,} \\ 
-\delta^{(4)}(x-y) \;\;\; & {\rm for} & \;\; x_0 \;,\; y_0 \;\; 
{\rm from \;\; the \;\; lower \;\; branch.} \end{array} \right. 
\end{displaymath}

Let us split the self-energy into three parts:
$$
\Pi (x,y) = \Pi_{\delta}(x)\delta^{(4)}(x,y) + \Pi^>(x,y) \Theta (x_0,y_0)
+ \Pi^<(x,y) \Theta (y_0,x_0) \;. 
$$
As we shall see later, $\Pi_{\delta}$ provides a dominant contribution
to the mean-field while $\Pi^{\gl}$ determines the collision terms of the
transport equations. 

With the help of the retarded and advanced Green functions 
(\ref{ret}) and the retarded and advanced self-energies defined in an 
analogous way, the equations (\ref{DS1}) and (\ref{DS2}) can be 
rewritten as  
\begin{eqnarray}\label{DSgl1}
\big[ \partial^2_x + m^2 &-& \Pi_{\delta}(x)\big ] 
\Delta^{\gl}(x,y) 
\nonumber \\ 
&=& \int d^4x' \Bigr[ \Pi^{\gl }(x,x') \Delta^{-}(x',y)+
\Pi^{+}(x,x') \Delta^{\gl}(x',y) \Bigl]  \;,
\end{eqnarray}
\begin{eqnarray}\label{DSgl2}
\big[ \partial^2_y  + m^2 &-& \Pi_{\delta}(y) \big] 
\Delta^{\gl }(x,y) 
\nonumber \\ 
&=&\int d^4x' \Bigr[ \Delta^{\gl}(x,x') \Pi^{-}(x',y)+
\Delta^{+}(x,x') \Pi^{\gl}(x',y)\Bigl]  \;, 
\end{eqnarray}
where all time integrations run from $- \infty$ to $+ \infty$.

\section{Towards the Transport Equation}

The transport equations are derived under the assumption that 
the Green functions and the self-energies depend weakly on the sum 
of their arguments and that they are significantly different from zero 
only when the difference of their arguments is close to zero. 
For homogeneous systems, the dependence on $X = (x+y)/2$ drops out entirely
due to the translational invariance and $\Delta (x,y)$ depends only on 
$u = x-y$. For weakly inhomogeneous, or quasihomogeneous systems, 
the Green functions and self-energies are assumed to vary slowly with
$X$. We additionally assume that the Green functions and self-energies 
are strongly {\it peaked} near $u = 0$, which means that the
correlation length is {\it short}.

We will now convert the equations (\ref{DSgl1}, \ref{DSgl2}) into the
transport and mass-shell equations by implementing the above approximation 
and performing the Wigner transformation (\ref{Wigner}) for all Green 
functions and self-energies. This is done by means of the translation 
rules such as: 
\begin{eqnarray*}
\int d^4x' f(x,x') g(x', y) & \longrightarrow &
f(X,p)g(X,p) \\ 
& + & {i \over 2} \biggr[
{\partial f(X,p) \over \partial p_{\mu}}
{\partial g(X,p) \over \partial X^{\mu}} - 
{\partial f(X,p) \over \partial X^{\mu}}
{\partial g(X,p) \over \partial p_{\mu}} \biggl] \;, \\
h(x) g(x, y) & \longrightarrow &
h(X)g(X,p) -  {i \over 2} {\partial h(X) \over \partial X^{\mu}}
{\partial g(X,p) \over \partial p_{\mu}}\;, \\
\partial^{\mu}_x f(x,y) & \longrightarrow &
(-ip^{\mu} + {1 \over 2} \partial^{\mu})f(X,p) \;.
\end{eqnarray*}
Here $\partial^{\mu} \equiv {\partial \over \partial X_{\mu}}$ and the 
functions $f(x,y)$ and $g(x,y)$ satisfy the assumptions discussed above. 
The function $h(x)$ is assumed to be weakly dependent on $x$.

The kinetic theory deals only with averaged system characteristics.
Thus, one usually assumes that the system is homogeneous on a scale 
of the Compton wave length of the quasiparticles. In other words, 
the characteristic length of inhomogeneities is assumed to be much 
larger than the inverse mass of quasiparticles. Therefore, we impose 
the condition
\begin{equation}\label{quasipar}
\Big\vert \Delta^{\gl }(X,p) \Big\vert \gg 
\Big\vert {1 \over m^2}\partial^2 
\Delta^{\gl}(X,p) \Big\vert \;,
\end{equation}
which leads to the quasiparticle approximation. The requirement 
(\ref{quasipar}) renders the off-shell contributions to the Green functions 
$\Delta^{\gl}$ negligible. Thus, we deal with the quasiparticles having 
on-mass-shell momenta.

Applying the translation rules and the quasiparticle approximation
to Eqs.~(\ref{DSgl1}, \ref{DSgl2}), we obtain
\begin{eqnarray}\label{trans}
\Bigr[p^{\mu} \partial_{\mu} - {1 \over 2} \partial_{\mu}
\Pi_{\delta}(X) \partial^{\mu}_p \Bigl] \Delta^{\gl }(X,p) 
& = & {i \over 2} \Big( \Pi^>(X,p) \Delta^<(X,p) -
      \Pi^< (X,p) \Delta^> (X,p) \Big) \nonumber \\
& - & {1 \over 4} 
\Big\{ \Pi^{\gl}(X,p), \Delta^+(X,p) + \Delta^-(X,p) \Big\}  \nonumber \\
& - &  {1 \over 4} 
\Big\{ \Pi^+(X,p) + \Pi^-(X,p),\, \Delta^{\gl}(X,p) \Big\} \;,
\end{eqnarray}
\begin{eqnarray}\label{mass}
\Bigr[ -  p^2  &+&  m^2 - \Pi_{\delta}(X) \Bigl] 
\Delta^{\gl }(X,p)  \nonumber \\
& = & {1 \over 2} \Big( \Pi^{\gl }(X,p) 
\big( \Delta^{+}(X,p)+ \Delta^{-}(X,p) \big) 
      + \big( \Pi^{+}(X,p) + \Pi^{-}(X,p) \big) 
\Delta^{\gl }(X,p) \Big) \nonumber \\
& + & {i \over 4} \Big\{ \Pi^>(X,p),\,\Delta^<(X,p) \Big\} 
 -  {i \over 4} \Big\{ \Pi^<(X,p),\, \Delta^>(X,p) \Big\} \;,
\end{eqnarray}
where we have introduced the Poisson-like bracket defined as
$$
\Big\{ C(X,p),\, D(X,p) \Big\} \equiv
{\partial C(X,p) \over \partial p_{\mu}}
{\partial D(X,p) \over \partial X^{\mu}} - 
{\partial C(X,p) \over \partial X^{\mu}}
{\partial D(X,p) \over \partial p_{\mu}} \;.
$$

One recognizes Eq.~(\ref{trans}) as a transport equation while 
Eq.~(\ref{mass}) as a so-called mass-shell equation. We write 
down these equation in a more compact way: 
\begin{eqnarray}\label{trans1}
\Big\{ p^2 - m^2 + \Pi_{\delta}(X) &+& {\rm Re} \Pi^+(X,p),\, 
\Delta^{\gl }(X,p) \Big\}   \nonumber \\
& = & i \Big( \Pi^>(X,p) \Delta^<(X,p) -
      \Pi^< (X,p) \Delta^> (X,p) \Big) \nonumber \\
& - & \Big\{ \Pi^{\gl}(X,p),\, {\rm Re} \Delta^+(X,p) \Big\}  \;,
\end{eqnarray}
\begin{eqnarray}\label{mass1}
\Bigr[ p^2 - m^2 &+& \Pi_{\delta}(X) + {\rm Re} \Pi^{+}(X,p) 
\Bigl] \Delta^{\gl }(X,p)  
= - \Pi^{\gl }(X,p) {\rm Re} \Delta^{+}(X,p) 
\nonumber \\
&-& {i \over 4} \Big\{ \Pi^>(X,p),\,\Delta^<(X,p) \Big\} 
 +  {i \over 4} \Big\{ \Pi^<(X,p),\, \Delta^>(X,p) \Big\} \;.
\end{eqnarray}
The gradient terms in the right-hand-sides of 
Eqs.~(\ref{trans1}, \ref{mass1}) are usually neglected \cite{Mro90,Mro94a}. 

We introduce the spectral function $A$ defined as
$$
A(x,y) \buildrel \rm def \over = \langle [ \phi (x),  \phi (y)] \rangle 
= i\Delta^>(x,y)  -  i\Delta^<(x,y) \;,
$$
where $[\phi (x),  \phi (y)]$  denotes the field commutator. Due to the 
equal time commutation relations
$$
[ \phi (t,{\bf x}),  \phi (t,{\bf y})] = 0 \;,\;\;\;\;\;\;\;
[\dot \phi (t,{\bf x}), \phi (t,{\bf y})] = -i \delta^{(3)}
({\bf x}-{\bf y}) \;, 
$$
with the dot denoting the time derivative, the Wigner transformed
spectral function satisfies the two identities 
\begin{eqnarray*}
\int {dp_0 \over 2\pi } \,  A(X,p) = 0 \;, \;\;\;\;\;\;\;
\int {dp_0 \over 2\pi } \, p_0 \, A(X,p) =  1 \;.
\end{eqnarray*}

From the transport and mass-shell equations (\ref{trans1}, \ref{mass1}) 
one immediately finds the equations satisfied by $A(X,p)$ which are
\begin{equation}\label{trans-spec}
\Big\{ p^2 - m^2 + \Pi_{\delta}(X) + {\rm Re} \Pi^+(X,p), \,
A(X,p) \Big\} = 2\Big\{ {\rm Im}\Pi^+(X,p),\,{\rm Re}\Delta^+(X,p) \Big\} \;,
\end{equation}
\begin{equation}\label{mass-spec}
\Bigr[ p^2 - m^2 + \Pi_{\delta}(X) + {\rm Re} \Pi^+(X,p) 
\Bigl] A(X,p)  
= 2 \, {\rm Im}\Pi^+(X,p) \, {\rm Re} \Delta^+(X,p) \;.
\end{equation}
One solves the algebraic equation (\ref{mass-spec}) as
\begin{eqnarray}\label{spec}
A (X,p) = { 2 {\rm Im}\Pi^+(X,p)
\over \big( p^2  -  m^2 + \Pi_{\delta}(X) 
+ {\rm Re}\Pi^+(X,p) \big)^2 + \big({\rm Im}\Pi^+(X,p) \big)^2} \;.
\end{eqnarray}
Then, it is shown that the function of the form (\ref{spec}) solves
Eq.~(\ref{trans-spec}) as well. The spectral function of the free fields 
can be found as
\begin{eqnarray*}
A_0 (X,p) = 2\pi \delta (p^2 - m^2 )
\bigr( \Theta (p_0) - \Theta (-p_0) \bigl) \;.
\end{eqnarray*}
Since ${\rm Re}\Pi^+$ determines the quasiparticle effective mass and 
${\rm Im}\Pi^+$ its width, the spectral function characterises the 
quasiparticle properties.

\section{Perturbative expansion}

As discussed in e.g. \cite{Dan84,Cal88,Hen90} the contour 
Green functions admit a perturbative expansion very similar to that 
known from the vacuum field theory with essentially the same Feynman rules.
However, the time integrations do not run from $-\infty $ to $+\infty $, 
but along the contour shown in Fig.~1. The right turning point of the
contour ($t_{max}$) must be above the largest time argument of the evaluated
Green function. In practice, $t_0$ is shifted to $-\infty $ and $t_{max}$ 
to $+\infty $. The second difference is the appearance of tadpoles, 
i.e. loops formed by single lines, which give zero contribution 
in the vacuum case. A tadpole corresponds to a Green function with two 
equal space-time arguments. Since the Green function  $\Delta(x,y)$ is 
not well defined for $x = y$ we ascribe the function $-i\Delta^<(x,x)$ 
to each tadpole. The rest of Feynman rules can be taken from the textbook 
of Bjorken and Drell \cite{Bjo64}.

The lowest-order contribution to the self-energy, which is associated
with the graph from Fig.~2, equals
$$
\Pi (x,y) = - {ig\over 2} \delta^{(4)}(x,y) \Delta^<_0(x,x)  \;,
$$
giving
\begin{equation}\label{pi-delta}
\Pi_{\delta}(x) =  -{ig\over 2} \Delta^<_0(x,x) \;,
\end{equation}
and
$$
\Pi^>(x,y) = \Pi^<(x,y) = 0 \;.
$$

The one-particle irreducible $g^2$ contributions to the self-energy
are shown in Fig.~3. The contribution corresponding to the diagram 3a 
can be easily computed. However, it is pure real and the only effect 
of this contributions is a higher order modification of the mean-field
term. Thus, we do not consider this diagrams but instead we analyse 
the contribution 3b which provides a qualitatively new effect. It
gives the contour self-energy equal
$$
\Pi_c(x,y) = {g^2 \over 6} \Delta_0(x,y) \Delta_0(y,x) \Delta_0(x,y) \;,
$$
and consequently
\begin{equation}\label{self-g2}
\Pi^{\gl}(x,y) = {g^2 \over 6}
\Delta^{\gl}_0(x,y) \Delta^{\lg}_0(y,x) \Delta^{\gl}_0(x,y)\;. 
\end{equation}

\section{Distribution function and transport equation}

The distribution function $f(X,p)$ is defined as 
\begin{eqnarray*}
\Theta (p_0) i \Delta^<(X,p) = \Theta (p_0)\; A(X,p) \; f(X,p) \;,
\end{eqnarray*}
where $A(X,p)$ is the spectral function (\ref{spec}). Then, one finds
\cite{Mro97a} that
\begin{eqnarray}\label{gr-f}
i \Delta^>(X,p)  =  \Theta (p_0) \; A(X,p) \; \big( f(X,p) + 1 \big)
- \Theta (-p_0) \; A(X,p) \; f(X,-p) \;,
\end{eqnarray}
\begin{eqnarray}\label{sm-f}
i \Delta^<(X,p)  = \Theta (p_0) \; A(X,p) \; f(X,p) 
- \Theta (-p_0) \; A(X,p) \; \big( f(X,-p) + 1 \big) \;.
\end{eqnarray}

There is a very important property of $\Delta^{\gl}$ expressed in the
from (\ref{gr-f}, \ref{sm-f}). Namely, if the Green functions 
$\Delta^{\gl}$ satisfy the transport equation (\ref{trans1}) and 
the spectral function solves the equation (\ref{mass-spec}), the 
mass-shell equation of $\Delta^{\gl}$, i.e. Eq.~(\ref{mass1}), is satisfied 
{\it automatically} in the 0-th order of the gradient expansion. 
Let us note that the quasiparticle dispersion relation is found 
as a solution of the equation
\begin{equation}\label{dispers}
p^2 - m^2 + \Pi_{\delta}(X) + {\rm Re} \Pi^+(X,p) = 0 \;.
\end{equation}

The distribution function $f$ satisfies the transport equation
which can be obtained from Eq.~(\ref{trans1}) for $\Delta^>$ or $\Delta^<$. 
After using Eq.~(\ref{trans-spec}) one finds
\begin{eqnarray}\label{trans-f}
A(X,p)\,\Big\{ p^2 - m^2 &+& {\rm Re}\Pi^+(X,p), 
\, f(X,p) \Big\} \nonumber \\
&=& i A(X,p) \,\Big( \Pi^>(X,p) \, f(X,p) -
      \Pi^< (X,p) \, \big(f(X,p)+1) \Big) \nonumber \\
&+& if(X,p) \, \Big\{ \Pi^>(X,p),\, {\rm Re} \Delta^+(X,p) \Big\} \nonumber \\
&-& i\big(f(X,p)+1\big)\,\Big\{ \Pi^<(X,p),\,{\rm Re} \Delta^+(X,p) \Big\}\;,
\end{eqnarray}
where $p_0 > 0$. We have also used here the following property of the 
Poisson-like brackets:
$$
\big\{A ,\, B\,C \big\} = \big\{A ,\, B \big\}\,C 
+ \big\{A ,\, C \big\}\,B \;.
$$

The left-hand-side of Eq.~(\ref{trans-f}) is a straightforward generalization
of the drift term of the standard relativistic transport equation. Computing
the Poisson-like bracket and imposing the mass-shell constraint one finds 
the familiar structure 
\begin{eqnarray*}
{1 \over 2} \Theta(p_0) \,
\Big\{ p^2 - m^2 &+& {\rm Re}\Pi^+(X,p), \, f(X,p) \Big\} \\
&=& E_p \Big({\partial \over \partial t} + {\bf v} \nabla \Big) f(X,p)
- \nabla {\rm Re}\Pi^+(X,p) \cdot \nabla_p f(X,p) \;,
\end{eqnarray*}
where the velocity ${\bf v}$ equals $\partial E_p /\partial {\bf p}$
with the (positive) energy $E_p$ being the solution of the dispersion
equation (\ref{dispers}).

Let us now analyse the right-hand-side of Eq.~(\ref{trans-f}). The 
collision terms are provided by the self energies (\ref{self-g2}).
Since the quasiparticles of interest are narrow 
($m^2 + {\rm Re}\Pi^+ \gg {\rm Im}\Pi^+$), we take into account only 
those terms contributing to $\Pi^{\gl}$ which are nonzero for the 
on-mass-shell momenta. The other terms are negligibly small \cite{Mro97a}. 
Then, the first term in r.h.s of the transport equation (\ref{trans-f}) 
is very similar to the standard collision term \cite{Gro80} of the 
Nordheim \cite{Nor28} (or Uehling-Uhlenbeck \cite{Ueh33}) form. Indeed,
\begin{eqnarray*}
i\Big( \Pi^>(X,p) \, f(X,p) &-& \Pi^< (X,p) \, \big(f(X,p)+1) \Big) \\ 
&=& {g^2\over 2} \int {d^4 k A^+_k \over (2\pi )^4} \,
{d^4 q A^+_q \over (2\pi )^4} \,
{d^4 r A^+_r \over (2\pi )^4} \, 
(2\pi)^4 \delta^{(4)} (p + q - k - r) \nonumber\\
&\times& \bigg( (f^p + 1) \;(f^q + 1) \, f^k  \, f^r - 
f^p\, f^q \, (f^k+1)  \, (f^r +1) \bigg)\,, \nonumber
\end{eqnarray*}
where $A^+_k \equiv \Theta(k_0) \; A(X,k)$ and $f^k \equiv f(X,k)$. The 
last two terms from r.h.s of Eq.~(\ref{trans-f}), which are neglected in 
the usual transport equation, are discussed in \cite{Mro97a}.

\chapter{Transport equations of quarks and gluons}

In this chapter we introduce the gauge dependent distribution functions 
of quarks and gluons. Then, we discuss the transport equations satisfied 
by these functions.  Finally, a very useful notion of the locally colorless 
plasma is considered.

\section{Distribution Functions}

The (anti-)quark distribution function $Q(p,x)\; 
\bigr( \bar Q(p,x)\bigl)$ is a hermitian $N_c\times N_c$ matrix in color 
space (for a $SU(N_c)$ color group) with $p$ denoting the quark four-momentum 
and $x$ the space-time coordinate \cite{Hei83,Win84,Elz86a}. The function 
transforms under local gauge transformations as 
\begin{equation}\label{2.1a}
Q(p,x) \rightarrow U(x)Q(p,x)U^{\dag }(x) \;.
\end{equation}
The color indices are here and in most cases below suppressed. 

The gluon distribution function \cite{Elz86b} is a hermitian 
$(N_c^2-1)\times (N_c^2-1)$ matrix \cite{Mro89} which transforms as
\begin{equation}\label{2.1b}
G(p,x) \rightarrow M(x)G(p,x)M^{\dag }(x) \;,
\end{equation}
where
$$
M_{ab}(x) = {\rm Tr}\bigr[\tau_a U(x) \tau_b U^{\dag }(x)]
$$
with $\tau_a ,\; a = 1,...,N_c^2-1$ being the $SU(N_c)$ group generators
in the fundamental representation. One sees that, in contrast to the 
distribution functions known from the physics of atomic gases, 
the distribution functions of quarks and gluons have no simple 
probabilistic interpretation due to the gauge dependence. This is, 
however, not surprising if one realizes that the question about the 
probability to find, let us say, a red quark in a phase-space cell 
centered around $(p,x)$ is not physical since the color of a quark can 
be changed by means of a gauge transformation. 

It follows from the transformation laws (\ref{2.1a}, \ref{2.1b}) that 
the traces of the distribution functions are gauge independent, 
and consequently they can have a familiar probabilistic interpretation. 
Indeed, the probability to find a quark of arbitrary color in a cell 
$(p,x)$ is of physical meaning since it is gauge independent.  
The quantities, which are color (gauge) independent like the baryon current 
$b^{\mu}$ or the energy momentum tensor $t^{\mu \nu}$, are entirely 
expressed through the traces of the distribution functions 
\begin{eqnarray*}
b^{\mu}(x) &=& {1 \over 3} \int {d^3p \over (2\pi )^3E} \; p^{\mu } 
\Big[ {\rm Tr}\big[Q(p,x)\big] - {\rm Tr}\big[\bar Q (p,x)\big]
\Big] \;, \\
t^{\mu \nu}(x) &=& \int {d^3p \over (2\pi )^3E} \; p^{\mu}p^{\nu} 
\Big[ {\rm Tr}\big[Q(p,x)\big] + {\rm Tr}\big[\bar Q (p,x)\big]
+{\rm Tr}\big[G(p,x)\big] \Big] \;, 
\end{eqnarray*}
with $E$ being the quark or gluon energy. Both quarks and gluons are 
assumed to be massless and their spin is treated as an internal degree 
of freedom.

The color current, which is a gauge dependent quantity, is expressed not 
only through the traces of the distribution functions but also through the
functions themselves. In the $N_c \times N_c$ matrix notation the current 
reads
\begin{eqnarray}\label{2.4}
j^{\mu }(x) = - {1 \over 2}g \int {d^3p \over (2\pi )^3E} \; p^{\mu } 
\Big[ Q(p,x) - \bar Q (p,x) &-& {1 \over N_c}{\rm Tr}\big[Q(p,x) - 
\bar Q (p,x)\big] \nonumber \\
&+&  2i \tau_a f_{abc} G_{bc}(p,x)\Big] \;,
\end{eqnarray}
where $g$ is the QCD coupling constant and $f_{abc}$ are the structure 
constants of the $SU(N_c)$ group. In the adjoint representation the 
color current (\ref{2.4}) is
\begin{eqnarray*}
j^{\mu }_a(x) = - g \int {d^3p \over (2\pi )^3E} \; p^{\mu } 
\Big[ {\rm Tr}\big[ \tau_a \big( Q(p,x) - \bar Q (p,x)\big) \bigl] 
+ i f_{abc} G_{bc}(p,x)\Big] \;,
\end{eqnarray*}
where we have used the equality 
${\rm Tr}(\tau_a \tau_b) = {1\over 2} \delta_{ab}$.

\section{Transport Equations}

The distribution functions of quarks and gluons satisfy the following
set of transport equations \cite{Hei83,Win84,Elz86a,Elz86b,Elz89,Mro89}:
\begin{eqnarray}\label{2.5}
p^{\mu} D_{\mu}Q(p,x) + g p^{\mu}{\partial \over \partial p_{\nu}}
{1 \over 2} \lbrace F_{\mu \nu}(x),Q(p,x)\rbrace &=& C[Q,\bar Q,G]\;, 
\nonumber \\
p^{\mu} D_{\mu}\bar Q(p,x) - g p^{\mu}{\partial \over \partial p_{\nu}}
{1 \over 2} \{ F_{\mu \nu}(x),\bar Q(p,x)\} &=& \bar C[Q,\bar Q,G]\;, 
\nonumber \\
p^{\mu} {\cal D}_{\mu}G(p,x) + g p^{\mu}
{\partial \over \partial p_{\nu}}
{1 \over 2} \{ {\cal F}_{\mu \nu}(x),G(p,x)\} &=& C_g[Q,\bar Q,G]\;, 
\end{eqnarray}
where $\{...,...\}$ denotes the anicommutator; $D_{\mu}$ and 
${\cal D}_{\mu}$ are the covariant derivatives which act as
$$
D_{\mu} = \partial_{\mu} - ig[A_{\mu}(x),...]\;,\;\;\;\;\;\;\;
{\cal D}_{\mu} = \partial_{\mu} - ig[{\cal A}_{\mu}(x),...]\;,
$$
where
$A_{\mu }$ and ${\cal A}_{\mu }$ are the mean-field four-potentials
defined as
$$
A^{\mu }(x) = A^{\mu }_a (x) \tau_a \;,\;\;\;\;\;
{\cal A}^{\mu }_{ab}(x) = - if_{abc}A^{\mu }_c (x) \;.
$$
$F_{\mu \nu}$ and ${\cal F}_{\mu \nu}$ are the mean-field stress
tensors with a color index structure analogous to that of the 
four-potentials. The mean-field is generated by the color current 
(\ref{2.4}) and the respective equation is
\begin{equation}\label{2.6}
D_{\mu} F^{\mu \nu}(x) = j^{\nu}(x)\;. 
\end{equation}
$C$, $\bar C$ and $C_g$ are the collision terms which vanish in the 
collisionless limit, i.e. when the plasma evolution is dominated by 
the mean-field effects\footnote{This occurs when the characteristic 
mean-field frequency is much greater than the parton collision frequency.}. 
As already mentioned, the collision terms of the QGP kinetic equations 
have not been systematically derived yet and their structure remains
obscure. The situation simplifies in the case of the colorless plasma
discussed in the next section. We note that the set of transport equations 
(\ref{2.5}, \ref{2.6}) is covariant with respect to the gauge 
transformations (\ref{2.1a}, \ref{2.1b}).

\section{Colorless Plasma}

Evolving towards thermodynamical equilibrium the system of quarks 
of gluons tends to neutralize color charges. It is expected
\cite{Sel94} that after a short period of time the plasma becomes 
locally colorless, the color current and the mean-field 
$F^{\mu \nu}$ vanish. Then, the distribution functions of quarks 
and gluons are proportional to the unit matrices in the color space. 
Specifically,
\begin{eqnarray*}
Q_{ij}(p,x) & = & {1 \over N_c}\, \delta_{ij}\, q(p,x) \;,\;\;\;
i,j = 1,...,N_c \;, 
\\
\bar Q_{ij}(p,x) & = & {1 \over N_c}\, \delta_{ij}\, \bar q(p,x) \;, 
\\
G_{ab}(p,x) & = & {1 \over N_c^2 - 1}\, \delta_{ab}\, g(p,x) \;,\;\;\;
a,b = 1,...,N_c^2-1 \;.
\end{eqnarray*}
As seen the distribution functions of the colorless plasma are gauge 
invariant.

The transport equations of the colorless plasma are essentially
simplified. Indeed, taking the trace of Eqs.~(\ref{2.5}) one finds 
\begin{eqnarray}\label{boltz}
p^{\mu} \partial_{\mu} q(x,p) & = & c [q,\bar q,g] \;,
\nonumber \\
p^{\mu} \partial_{\mu} \bar q(x,p) & = &  \bar c [q,\bar q,g] \;,
\nonumber \\
p^{\mu} \partial_{\mu} g(x,p) & = &   c_g [q,\bar q,g] \;,
\end{eqnarray}
where $c \equiv {\rm Tr}C$, $\bar c \equiv {\rm Tr}\bar C$ and
$c_g \equiv {\rm Tr}C_g$. Because the trace of a commutator is zero, 
the covariant derivatives reduce to the normal ones in (\ref{boltz}).

In the case of a colorless plasma the color charges can be treated as internal 
degrees of freedom of the quarks and gluons, and it is sufficient to operate 
with the color averaged quantities which are gauge independent. Then, one 
can imitate the dynamics of the colorless plasma with a non-gauge field 
theory model such as $\phi^4$. Then, the collision terms are of the Nordheim 
\cite{Nor28} (or Uehling-Uhlenbeck \cite{Ueh33}) form as discussed in
the previous chapter. Therefore, even not knowing the collision terms $C$, 
$\bar C$ and $C_g$, we expect that the respective terms of the colorless 
plasma  $c$, $\bar c$ and $c_g$, which represent the binary collisions, are 
\begin{eqnarray}\label{Nord}
\lefteqn{ c [q,\bar q,g] = \int {d^3p_2 \over (2\pi )^3E_2}
{d^3p_3 \over (2\pi )^3E_3} {d^3p_4 \over (2\pi )^3E_4}}
\nonumber \\
\Big[ &\frac{1}{2}& \big[ q_3 q_4 (1 - q_1) (1 - q_2) -
q_1 q_2 (1 - q_3) (1 - q_4) \big] 
W_{q q\rightarrow q q} (p_3,p_4 \vert p_1,p_2)
\nonumber \\
&+& \big[ q_3 \bar q_4 (1 - q_1) (1 - \bar q_2) - q_1 \bar q_2 (1 - q_3)
(1 - \bar q_4) \big] 
W_{q \bar q \rightarrow q \bar q} (p_3,p_4 \vert p_1,p_2)
\nonumber \\
&+& \big[ q_3 g_4 (1 - q_1) (1 + g_2) - q_1 g_2 (1 - q_3)
(1 + g_4) \big] W_{q g \rightarrow q g} (p_3,p_4 \vert p_1,p_2)
   \nonumber \\
&+& \big[ g_3 g_4 (1 - q_1) (1 - \bar q_2) 
- q_1 \bar q_2 (1 + g_3) (1 + g_4) \big] 
W_{q \bar q \rightarrow g g} (p_3,p_4 \vert p_1,p_2)
 \Big]   \nonumber \\
\end{eqnarray}
with the analogous expressions for $\bar c [\bar q,q,g]$ and 
$c_g [g,q,\bar q,]$. We have used here the abbreviations 
$q_1 \equiv q (x, p_1)$, $q_2 \equiv q (x, p_2)$  etc. Furthermore 
$p_1 \equiv  p$. The coefficient $\frac{1}{2}$ in the first line of 
the r.h.s. of Eq.~(\ref{Nord}) is required to avoid the double counting of 
identical particles. The quantities like 
$W_{q g \rightarrow q g} (p_3,p_4 \vert p_1,p_2)$, which corresponds 
to the quark-gluon scattering, are equal to the square of the respective 
matrix element multiplied by  the energy-momentum conserving 
$\delta -$ function. We note that the collision terms have to satisfy 
the relations 
\begin{eqnarray*}
\int {d^3p \over (2\pi )^3E} \Big[ c[q,\bar q,g] - 
\bar c[q,\bar q,g]\Big] &=& 0\;, \\ \nonumber
\int {d^3p \over (2\pi )^3E} p^{\mu} \Big[ c[q,\bar q,g] +  
\bar c[q,\bar q,g] + c_g[q,\bar q,g] \Big] &=& 0 \;,
\end{eqnarray*}
in order to be consistent with the baryon number and energy-momentum 
conservation.

In the variety of applications one uses the collision terms in the 
relaxation time approximation i.e.
\begin{eqnarray}\label{trans-relax}
c &=& \nu p_{\mu}u^{\mu}(x)\Bigr(q^{eq}(p,x) - q(p,x)\Bigl)
\;,\\ \nonumber
\bar c &=& \bar \nu p_{\mu}u^{\mu}(x)\Bigr(\bar q^{eq}(p,x) -\bar q(p,x)\Bigl)
\;, \\ \nonumber
c_g &=& \nu_g p_{\mu}u^{\mu}(x) \Bigr(g^{eq}(p,x) - g(p,x)
\Bigl)\;,
\end{eqnarray}
where $\nu$,  $\bar \nu$ and $\nu_g$ are the collision frequencies and
$u^{\mu}$ is the hydrodynamic four-velocity which defines the local rest
frame of the quark-gluon system. The equilibrium distribution functions
are
\begin{eqnarray*}
q^{eq}(p,x)&=& {2N_f N_c \over \exp \bigr(\beta^{\mu}(x)p_{\mu} - 
\beta (x)\mu (x) \bigl)+1} \;, \\ \nonumber
\bar q^{eq}(p,x)&=& {2N_f N_c \over \exp \bigr(\beta^{\mu}(x)p_{\mu} + 
\beta (x)\mu (x) \bigl)+1}\;, \\ \nonumber
g^{eq}(p,x)&=& {2(N_c^2 -1) \over\exp \bigr(\beta^{\mu}(x)p_{\mu} \bigl)- 
1}\;, 
\end{eqnarray*}
where $\beta^{\mu}(x)\equiv\beta (x) u^{\mu}(x)$, $\beta (x) \equiv
T^{-1}(x)$; $T(x)$ and $\mu (x)$ are the local temperature and quark 
chemical potential, respectively; $N_f$ is the number of quark flavours.
Spin, flavour and color are treated here as internal degrees of freedom.

\chapter{Plasma color response}

In this chapter we discuss how the plasma, which is colorless,
homogeneous and stationary, responses to the color small fluctuations.

\section{Linear response analysis}

The distribution functions are assumed to be of the form
\begin{eqnarray}\label{5.1}
Q_{ij}(p,x) &=& n(p)\delta_{ij} + \delta Q_{ij}(p,x) \;, 
\\ \nonumber
\bar Q_{ij}(p,x) &=& \bar n(p)\delta_{ij} + \delta \bar Q_{ij}(p,x) \;,
\\ \nonumber
G_{ab}(p,x) &=& n_g(p)\delta_{ab} + \delta G_{ab}(p,x) \;,
\end{eqnarray}
where the functions describing the deviation from the colorless state 
are assumed to be much smaller than the respective colorless functions.
The same is assumed for the momentum gradients of these functions.

Substituting (\ref{5.1}) in (\ref{2.4}) one gets
\begin{eqnarray}\label{5.2}
j^{\mu }(x) = &-& {1 \over 2}g \int {d^3p \over (2\pi )^3E}\, p^{\mu} 
\Bigr[ \delta Q(p,x) - \delta \bar Q (p,x) \\ \nonumber
&-& {1 \over N_c}Tr\bigr[\delta Q(p,x) - \delta \bar Q (p,x)\bigl]
+ 2i \tau_a f_{abc} \delta G_{bc}(p,x)\Bigl] \;.
\end{eqnarray}
As seen the current occurs due to the deviation of the system 
from the colorless state. When the system becomes neutral there 
is no current and one expects that there is no mean field. Therefore, 
we linearize Eq. (\ref{2.6}) with respect to the four potential 
to the form
\begin{eqnarray*}
\partial_{\mu} F^{\mu \nu}(x) = j^{\nu}(x)\; 
\end{eqnarray*}
with $F^{\mu \nu} = \partial^{\mu} A^{\nu} - \partial^{\nu} A^{\mu}$.
It should be stressed here that the linearization procedure does not
cancel all non-Abelian effects. The gluon-gluon coupling, which is of 
essentially non-Abelian character is included because the gluons contribute 
to the color current (\ref{5.2}). Let us also observe that in the linearized 
theory the color current is conserved (due to antisymmetry of $F^{\mu \nu}$) 
i.e. $\partial_{\mu} j^{\mu} = 0$. Finally we note that, as shown in
\cite{Bla94}, the semiclassical QCD transport theory effectively incorporates 
the resummation over the so-called hard thermal loops \cite{Bra90}. 

Now we substitute the distribution functions (\ref{5.1}) to the transport
equations (\ref{2.5}) with the collision terms (\ref{trans-relax}). 
Linearizing the equations with respect to $\delta Q$, $\delta \bar Q$ and 
$\delta G$, one gets
\begin{eqnarray}\label{5.4}
\Bigr( p^{\mu} \partial_{\mu} + \nu p_{\mu}u^{\mu} \Bigl)\delta Q(p,x) 
&=& - g p^{\mu}F_{\mu \nu}(x){\partial n(p) \over \partial p_{\nu}}
+ \nu p_{\mu}u^{\mu}\bigr( n^{eq}(p) - n(p)\Bigl) \;, 
\\ \nonumber
\Bigr( p^{\mu} \partial_{\mu} + \bar \nu p_{\mu}u^{\mu} \Bigl)
\delta \bar Q(p,x) 
&=& \;\;\; g p^{\mu}F_{\mu \nu}(x){\partial \bar n(p) \over \partial p_{\nu}}
+\bar \nu p_{\mu}u^{\mu}\bigr( \bar n^{eq}(p) - \bar n(p)\Bigl) \;, 
\\ \nonumber
\Bigr( p^{\mu} \partial_{\mu} + \nu_g p_{\mu}u^{\mu} \Bigl)\delta G(p,x) 
&=&- g p^{\mu}{\cal F}_{\mu \nu}(x){\partial n_g(p) \over \partial p_{\nu}}
+ \nu_g p_{\mu}u^{\mu}\bigr( n^{eq}_g(p) - n_g(p)\Bigl) \;.
\end{eqnarray}
Performing the linearization one should remember that $A^{\mu}$ is of
order of $\delta Q$. Treating the chromodynamic field as an external one, 
Eqs. (\ref{5.4}) are easily solved 
\begin{eqnarray}\label{5.5}
\delta Q(p,x) &=& - g \int d^4 x^{\prime}
\Delta_p(x - x^{\prime}) \Bigr[
p^{\mu}F_{\mu \nu}(x^{\prime}){\partial n(p) \over \partial p_{\nu}}
- \nu p_{\mu}u^{\mu}\bigr( n^{eq}(p) - n(p)\Bigl) \Bigl]\;, 
\\ \nonumber
\delta \bar Q(p,x) &=& \;\;\; g \int d^4 x^{\prime}
\Delta_p(x - x^{\prime}) \Bigr[
 p^{\mu}F_{\mu \nu}(x^{\prime}){\partial \bar n(p) \over \partial p_{\nu}}
+ \bar \nu p_{\mu}u^{\mu}\bigr( \bar n^{eq}(p) - \bar n(p)\Bigl) 
\Bigl]\;, 
\\ \nonumber
\delta G(p,x) &=& - g \int d^4 x^{\prime}
\Delta_p(x - x^{\prime}) \Bigr[
p^{\mu}{\cal F}_{\mu \nu}(x^{\prime}){\partial n_g(p) \over \partial p_{\nu}}
- \nu_g p_{\mu}u^{\mu}\bigr( n^{eq}_g(p) - n_g(p)\Bigl) \Bigl]\;, 
\end{eqnarray}
where $\Delta_p(x)$ is the Green function of the kinetic operator
with the collision term in the relaxation time approximation,
$$
\Delta_p(x) = E^{-1} \Theta (t) \, e^{-\nu^{\prime}t}\, \delta^{(3)}
({\bf x} - {\bf v}t) \;,
$$
with $t$ being the zero component of $x$, $x^{\mu} \equiv (t, {\bf x})$,
${\bf v} \equiv  {\bf p}/E $ and $\nu^{\prime} \equiv \nu p^{\mu}u_{\mu}$;
in the plasma rest frame $\nu^{\prime} = \nu $.

Substituting the solutions (\ref{5.5}) in Eq. (\ref{5.2}) and performing 
the Fourier transformation with respect to $x$-variable we get
\begin{equation}\label{5.6}
j^{\mu}(k) = \sigma^{\mu \rho \lambda}(k) F_{\rho \lambda}(k) 
\end{equation}
with the color conductivity tensor expressed as
\begin{eqnarray}\label{5.7}
\sigma^{\mu \rho \lambda}(k) 
&=& i {g^2 \over 2} \int {d^3p \over (2\pi )^3E} 
\Bigr[ {p^{\mu}p^{\rho} \over p^{\sigma}(k_{\sigma} + i\nu u_{\sigma})}
{\partial n(p) \over \partial p_{\lambda}} \\[2mm] \nonumber
&+&  {p^{\mu}p^{\rho} \over p^{\sigma}(k_{\sigma} + i\bar \nu u_{\sigma})}
{\partial \bar n(p) \over \partial p_{\lambda}} + 
{2N_c p^{\mu}p^{\rho} \over  p^{\sigma}(k_{\sigma} + i\nu_g u_{\sigma})}
{\partial n_g (p) \over \partial p_{\lambda}} \Bigl] \;.
\end{eqnarray}
If the plasma colorless state is isotropic, which is the case of the global 
equilibrium, one finds that 
$\sigma^{\mu \rho \lambda}(k) = \sigma^{\mu \rho}(k)u^{\lambda}$ and 
Eq. (\ref{5.6}) gets more familiar form of the Ohm law, which in the plasma 
rest frame reads
\begin{eqnarray*}
j^{\alpha}(k) = \sigma^{\alpha \beta}(k)E^{\beta}(k) \;,
\end{eqnarray*}
where the indices $\alpha ,\beta ,\gamma = 1,2,3$ label the space
axes and $E^{\alpha}(k)$ is the $\alpha$-component of the chromoelectric
vector. The conductivity tensor describes the response of the QGP to the
chromodynamic field. Within the approximation used here it is a color scalar
(no color indices) or equivalently is proportional to the unit matrix in the 
color space. In the next sections we will extract the information about QGP 
contained in $\sigma^{\mu \rho \lambda}(k)$.

\section{Chromoelectric permeability}

Let us introduce, as in the electrodynamics, the polarization vector 
${\bf P}(x)$ defined as 
\begin{equation}\label{6.1}
{\rm div}{\bf P}(x) = - \rho (x)\;,\;\;\;
{\partial \over \partial t}{\bf P}(x) = {\bf j}(x) \;,
\end{equation}
where $\rho $ and {\bf j} are the time-like and space-like components,
respectively, of the color induced four-current, $j^{\mu} = (\rho, {\bf j})$.
The definition (\ref{6.1}) is self-consistent, only when the color current is
conserved, not covariantly conserved. This just the case of the
linear response approach. Further, we define the chromoelectric induction 
vector ${\bf D}(x)$,
\begin{equation}\label{6.2}
{\bf D}(x) = {\bf E}(x) + {\bf P}(x)
\end{equation}
and the chromoelectric permeability tensor, which relates the
Fourier transformed {\bf D} and {\bf E} fields,
\begin{equation}\label{6.3}
D^{\alpha}(k) = \epsilon^{\alpha \beta}(k)E^{\beta}(k) \;,
\end{equation}
where $\alpha , \beta = 1, 2, 3$. Since the conductivity tensor (\ref{5.7}) 
is a color scalar the permeability tensor is a color scalar as well.

Using the definitions (\ref{6.1}, \ref{6.2}, \ref{6.3}) one easily finds that
\begin{equation}\label{6.4}
\epsilon^{\alpha \beta}(k) = \delta^{\alpha \beta}
 - {i \over \omega}\sigma^{\alpha 0 \beta}(k) - {i \over \omega^2}
\Bigr[k^{\gamma}\sigma^{\alpha \beta \gamma}(k) - 
k^{\gamma}\sigma^{\alpha \gamma \beta}(k) \Bigl]
\end{equation}
with $\sigma^{\alpha \gamma \beta}(k)$ given by Eq. (\ref{5.7}); 
$\omega $ is the time-like component of the wave four-vector, 
$k^{\mu}\equiv (\omega, {\bf k})$. For the isotropic plasma the two last 
terms in Eq. (\ref{6.4}) vanish. Substituting the conductivity tensor 
(\ref{5.7}) into Eq. (\ref{6.4}) we get the permeability tensor in the 
plasma rest frame 
\begin{eqnarray}
\epsilon^{\alpha \beta}(k) = \delta^{\alpha \beta}
&+& {g^2 \over 2\omega } \int {d^3p \over (2\pi )^3}
\Bigr[ {v^{\alpha} \over \omega - {\bf kv} + i\nu}
{\partial n(p) \over \partial p_{\gamma}} 
+ {v^{\alpha} \over \omega - {\bf kv} + i\bar \nu}
{\partial \bar n(p) \over \partial p_{\gamma}} \\[2mm] \nonumber
&+& 2 N_c {v^{\alpha} \over \omega - {\bf kv} + i\nu_g}
{\partial  n_g(p) \over \partial p_{\gamma}}
\Bigl] \Bigr[ \Bigr(1 - {{\bf kv} \over \omega} \Bigl) 
\delta^{\gamma \beta} + {k^{\gamma}v^{\beta}\over \omega} \Bigl] \;.
\label{epsilon}
\end{eqnarray}
In the case of the isotropic plasma the permeability tensor can be
expressed as
\begin{eqnarray*}
\epsilon^{\alpha \beta}(k) = \epsilon_T(k)
\Bigr( \delta^{\alpha \beta} - k^{\alpha}k^{\beta}/{\bf k}^2 \Bigl)
+\epsilon_L(k)\; k^{\alpha}k^{\beta}/{\bf k}^2  \;.
\end{eqnarray*}
with the longitudinal and transversal permeability functions equal
\begin{eqnarray}
\epsilon_L(k) = 1 &+& {g^2 \over 2\omega {\bf k}^2} 
\int {d^3p \over (2\pi )^3} 
\Big[ {{\bf kv}\,k^{\gamma}\over \omega - {\bf kv} + i\nu}
{\partial n(p) \over \partial p_{\gamma}} \\[2mm] \nonumber
&+& {{\bf kv}\,k^{\gamma}\over \omega - {\bf kv} + i\bar \nu}
{\partial \bar n(p) \over \partial p_{\gamma}} 
+ {{\bf kv}\,k^{\gamma}\over \omega - {\bf kv} + i\nu_g}
{\partial n_g(p) \over \partial p_{\gamma}}\Big]  \\[4mm] \label{epsL}
\epsilon_T(k) = 1 &+& {g^2 \over 2\omega} 
\int {d^3p \over (2\pi )^3} 
\Big[ {1\over \omega - {\bf kv} + i\nu}
{\partial n(p) \over \partial p_{\gamma}} \\[2mm] \nonumber
&+& {1\over \omega - {\bf kv} + i\bar \nu}
{\partial \bar n(p) \over \partial p_{\gamma}} 
+ {1\over \omega - {\bf kv} + i\nu_g}
{\partial n_g(p) \over \partial p_{\gamma}}\Big] 
\Big[v^{\gamma}- {{\bf kv}\,k^{\gamma} \over {\bf k}^2} \Big] \;. \label{epsT}
\end{eqnarray}

Because the QCD equations within the linear response approach coincide (up 
to the trivial matrix structure) with those of the electrodynamics, the 
dispersion relations of the plasma oscillations, or of plasmons, are those 
of the electrodynamics and they read \cite{Lan60,Sil61}
\begin{equation}\label{disp-eq-det}
\det \mid {\bf k}^2\delta^{\alpha \beta} - k^{\alpha}k^{\beta} 
- \omega^2\epsilon^{\alpha \beta}(k) \mid = 0 \;.
\end{equation}
The relation (\ref{disp-eq-det}) gets simpler form for the isotropic plasma.
Namely, 
\begin{equation}\label{6.6}
\epsilon_L (k) = 0\;,\;\;\;
\epsilon_T (k) = {\bf k}^2/\omega^2 \;.
\end{equation}
The dispersion relations determine the plasma waves which can be propagate
in the medium. Specifically, the plane wave with $\omega ({\bf k})$, which 
satisfies the dispersion equation (\ref{disp-eq-det}), automatically solves 
the sourceless Maxwell equations in a medium. Using the `quantum' language, 
the dispersion equation gives the relation between the energy and momentum 
of the quasiparticle excitations. In the case of plasma these are the 
transverse and longitudinal plasmons.

There are three classes of the solutions of Eq. (\ref{disp-eq-det}). Those 
with pure real $\omega $ are stable - the wave amplitude is constant in time.
If the imaginary part of frequency is negative, the oscillations are damped 
- the amplitude decreases in time. Of particular interest are the solutions 
with the positive Im$\omega$ corresponding to the so-called plasma 
instabilities - the oscillations with the amplitude exponentially growing 
in time.

The permeability tensor in the static limit ($\omega \rightarrow 0$)
provides the information about the plasma response to constant fields.
Computing $\epsilon_L (\omega =0,{\bf k})$ for the equilibrium collisionless
plasma one finds 
\begin{eqnarray*}
\epsilon_L (\omega =0,{\bf k}) = 1 +{m_D^2 \over {\bf k}^2} \;,
\end{eqnarray*}
where $m_D$ is the Debye mass which for the baryonless plasma of massless 
quarks and gluons equals
\begin{equation}\label{7.13}
m^2_D = {g^2 T^2 (N_f +2N_c) \over 6} \;.
\end{equation}
The chromoelectric potential of the static point-like source embeded in 
the plasma, which equals \cite{Lif81,Sil61}
$$
A_0({\bf x}) = {g \over 4\pi \mid {\bf x} \mid } 
\exp (-m_D\mid {\bf x}\mid ) \;,
$$
is screened at the distance $1/m_D$.

Since the parton density is $\sim T^3$, one finds from Eq. (\ref{7.13})
that the number of partons in the Debye sphere (the sphere of the radius 
equal to the screening length) is $\sim 1/g^3$. It is much greater than 
unity if the plasma is {\it perturbative} i.e. when $1/g \gg 1$. A large 
parton number in the Debye sphere justifies, in particular, the use of the 
mean-field to describe QGP. Let us also mention that the ultrarelativistic 
{\it perturbative} plasma is automatically {\it ideal} i.e. the average 
parton interaction energy, which is  $\sim g^2 / \langle r \rangle $ with 
$\langle r \rangle \sim T^{-1}$ being the average interparticle distance, 
is much smaller than the parton thermal energy which equals $\sim T$.
This is not the case for the nonrelativistic electron plasma. Then, the 
screening length is (see e.g. \cite{Lif81,Sil61})
$$
m^2_D = e^2 {n_e \over T} \;, 
$$
with $n_e$ being the electron density\footnote{We use the units, where the 
fine structure constant $\alpha = e^2/4\pi$. In the Gauss units, which are 
traditionally used the electron-ion plasma physics, $\alpha = e^2$ .},
which is independent of the temperature. As seen, the smallness of the 
coupling constant does not guarantee that the nonrelativistic plasma is 
ideal. This occurs when the number of electrons in the Debye sphere is 
large, i.e. when $T^{3/2} \gg e^3 n_e^{1/2}$.

\section{Oscillations around the global equilibrium}

Substituting the equilibrium distribution functions, Fermi-Dirac for 
quarks and Bose-Einstein for gluons, into Eqs. (\ref{epsL}) and 
(\ref{epsT}) one finds the permeability functions
$\epsilon_L$ and $\epsilon_T$, which for the collisionless 
($\nu = \bar \nu = \nu_g =0$) and baryonless ($\mu = 0$) plasma 
of massless partons can be computed analitically as
\begin{equation}\label{7.2}
\epsilon_L = 1 + {3\omega_0^2 \over k^2} \Biggr[ 1 - {\omega \over
2k} \Bigr[ \ln \mid {k+\omega \over k-\omega} \mid - i\pi
\Theta(k-\omega ) \Bigl] \Biggl] \;,
\end{equation}
\begin{equation}\label{7.3}
\epsilon_T = 1 - {3\omega_0^2 \over 2k^2} \Biggr[ 1 - \Bigr({\omega \over
2k}-{k \over 2\omega}\Bigl) \Bigr[ \ln \mid {k+\omega \over k-\omega} \mid 
- i\pi \Theta(k-\omega ) \Bigl] \Biggl] \;,
\end{equation}
where $k \equiv \mid {\bf k}\mid $ and  $\omega_0$ is the plasma frequency 
equal
\begin{equation}\label{7.4}
\omega_0^2 = {g^2 T^2 (N_f +2N_c) \over 18} \;.
\end{equation}
One sees that for $\omega > k$ the dielectric functions (\ref{7.2}, \ref{7.3})
are purely real i.e. there are no dissipative processes.

Substituting (\ref{7.2}, \ref{7.3}) into (\ref{6.6}) one finds the dispersion 
relation of the longitudinal mode (the chromoelectric field parallel to the 
wave vector) 
\begin{displaymath}
\omega^2 = \left\{ 
\begin{array}{ccl}
\omega^2_0 +{3 \over 5} k^2
\;, &{\rm for}&  \omega_0 \gg k \\  
k^2 \Bigr(1 + 4\exp (-2 -2k^2/3\omega^2_0)\Bigl)
\;, &{\rm for}& \omega_0 \ll k
\end{array}   \right.
\end{displaymath}
and 
\begin{displaymath}
\omega^2 = \left\{ 
\begin{array}{ccl}
\omega^2_0 +{6 \over 5} k^2
\;, &{\rm for}&  \omega_0 \gg k \\  
{3 \over 2}\omega^2_0 + k^2
\;, &{\rm for}& \omega_0 \ll k
\end{array}   \right.
\end{displaymath}
for the transverse one (the chromoelectric field perpendicular to 
the wave vector). Because the longitudinal and transverse 
oscillations are time-like ($\omega^2 > k^2$), the phase velocity 
of the waves is greater than the velocity of light. For this reason the 
Landau damping is absent. As known, the Landau damping is due to the 
collisionless energy transfer from the wave to the plasma particles, 
the velocity of which is equal to the wave phase velocity \cite{Lif81}. 

The oscillations of the collisionless QGP around global equilibrium 
have been studied by means of the transport theory in several papers 
\cite{Mro87a,Elz87,Bia88a,Bia88b,Mro89}. The problem has been also 
discussed in \cite{Hei85b,Hei86} using a specific variant of the QGP 
theory with the classical color \cite{Hei83,Hei85a}. In the above 
presentation we have followed \cite{Mro89}. The dispersion relations 
given above agree with those found in the finite-temperature QCD within 
the one-loop approximation, see e.g. \cite{Wel82,Kal84,Han87,Hei87}.

Let us now consider the dielectric function with nonzero equilibration
rates. As previously, the partons are massless and the plasma is baryonless 
which imposes $\nu = \bar \nu$. Then, one easily evaluates the integrals 
(\ref{epsL}) and (\ref{epsT}) for $\omega \gg  k $, $\omega \gg \nu $ 
and $\omega \gg \nu_g $. The results read \cite{Mro89}:
\begin{eqnarray*}
\omega^2 = \omega^2_0 - \zeta^2 + {3 \over 4}\phi^2 +
{3 \over 5} k^2 \;,\;\;\;\;\;\; \gamma = {1 \over 2} \phi \;
\end{eqnarray*}
for the longitudinal mode and
\begin{eqnarray*}
\omega^2 = \omega^2_0 - \zeta^2 + {3 \over 4}\phi^2 +
{6 \over 5} k^2 \;,\;\;\;\;\;\; \gamma = {1 \over 2} \phi \;,
\end{eqnarray*}
for the transverse one; $\omega $ and $\gamma $ denote the real and 
imaginary part, respectively, of the complex frequency; $\phi $ and 
$\zeta $ are parameters related to the equilibration rates,
\begin{eqnarray}\label{7.9}
\phi = \nu {N_f \over N_f + 2N_c} +\nu_g{2N_c \over N_f + 2N_c} \;, 
\\ \nonumber
\zeta^2= \nu^2{N_f \over N_f +2N_c} +\nu_g^2{2N_c \over N_f + 2N_c} \;.
\end{eqnarray}
One sees that, when compared with the collisionless plasma, the frequency 
of the oscillations is smaller and that the oscillations are damped. 
To find the numerical value of the damping rate - the plasma oscillation 
decrement $\gamma $, the equilibration rates ($\nu $ and $\nu_g $) have 
to be estimated.  

If $\nu$ or $\nu_g$ is identified with the mean free flight time controlled
by the binary collisions, the equilibration rate is of the order $g^4\ln 1/g$. 
However, in the relativistic plasma there is another damping mechanism
which is the plasmon decay into quark-antiquark or gluon-gluon pair.
It is easy to observe that, even in the limit of massless quarks, the 
decay into gluons is much more probable than that into quarks 
\cite{Hei87,Mro89}. Let us consider the decay of plasmon of zero momentum. 
The phase-space volume of the decay final state is proportional to  
\begin{eqnarray}\label{7.10}
\Bigr( 1 \mp n^{eq}(\omega_0/2) \Bigl)^2 \;, 
\end{eqnarray}
where the upper sign refers to quarks while the lower one to gluons. 
Since the plasma frequency (\ref{7.4}) is much smaller than the temperature 
in the perturbative plasma, the factor (\ref{7.10}) can be
expanded as
\begin{eqnarray*}
\Bigr( 1 - n^{eq}(\omega_0/2) \Bigl)^2 &\cong& 1/4 + \omega_0/8T \;, 
\\
\Bigr( 1 + n^{eq}(\omega_0/2) \Bigl)^2 &\cong& 4T^2/\omega_0^2 \;.
\end{eqnarray*}
One sees that the decay into gluons is more probable than that
into quarks by a factor of order $g^{-2}$ \cite{Hei87}.

Using the standard rules of finite-temperature field theory, one
easily finds (see e.g. \cite{Mro89}) the width of the zero-momentum 
plasmon due to the decay into gluons 
$$
\Gamma_d= {g^2N_c \over 2^43\pi}\,\omega_0 \Bigr( 1 + n^{eq}(\omega_0/2) 
\Bigl)^2 \cong {gN_cT \over 2^{3/2}\pi (N_f + 2N_c)^{1/2}} \;, 
$$
which is the same for longitudinal and transverse plasmons. However, 
$\Gamma_d$ cannot be identified with the plasmon equilibration rate
$\Gamma $, because the plasmon decays are partially compensated by the
plasmon formation processes. As shown in \cite{Wel83}, see also \cite{Hei87},
the formation rate $\Gamma_f$ is related to $\Gamma_d$ as
$$
\Gamma_f= \exp(-\omega_0/T) \Gamma_d \cong 
(1-\omega_0/T) \Gamma_d \;.
$$
Since the equilibration rate of the plasmon $\Gamma =\Gamma_d-\Gamma_f$ 
\cite{Wel83}, one finds \cite{Hei87}
\begin{equation}\label{7.11}
\Gamma \cong {g^2 N_cT \over 12 \pi } \;. 
\end{equation}
We note that $\Gamma_d$ and $\Gamma_f$ are of the order of $g$, while 
$\Gamma $ is of the order of $g^2$. Since there is a preferred reference 
frame - the rest frame of the thermostat, the plasmon decay width is not 
a Lorentz scalar. Therefore, the result (\ref{7.11}) is valid only for 
the zero-momentum or approximately long-wave plasmons.

Substituting $\nu_g$ equal (\ref{7.11}) and $\nu = 0$ into Eq. (\ref{7.9}), 
one estimates the decrement of the oscillation damping as
\begin{equation}\label{7.12}
\gamma \cong {g^2 \over 12\pi} {N_c^2 \over N_f + 2N_c}\;T \;. 
\end{equation}
Although $\nu =0$ the damping rate depends on the number of quark flavours. 
This seems to be in agreement with the physical intuition. When the number of
quark flavours is increased the inertia of the system is also
increased, and consequently the time needed to damp the oscillations
is longer. However Eq. (\ref{7.12}) disagrees (by a factor $2N_c/(N_f+2N_c)$) 
with the result from \cite{Hei87}, where $\gamma $ equals (\ref{7.11}). 
Unfortunately, the discrepancy can not be resolved within the relaxation
time approximation and a more elaborated analysis is needed.

\chapter{Filamentation instability}

In the near future the nucleus-nucleus collisions will be studied 
experimentally at the accelerators of a new generation: Relativistic 
Heavy-Ion Collider (RHIC) at Brookhaven and Large Hadron Collider (LHC) 
at CERN. The collision energy will be larger by one or two orders 
of magnitude than that one of the currently operating machines.  A copious 
production of partons, mainly gluons, due to hard and semihard processes 
is expected in the heavy-ion collisions at this new energy domain 
\cite{Gei95,Wan97}. Thus, one deals with the many-parton system at the early
stage of the collision. The system is on average locally colorless
but random fluctuations can break the neutrality. Since the system is 
initially far from equilibrium, specific color fluctuations can exponentially
grow in time and then noticeably influence the system evolution. While the 
very existence of such instabilities, similar to those which are known from 
the electron-ion plasma, see e.g. \cite{Akh75}, is fairly obvious and was 
commented upon long time ago \cite{Hei84}, it is far less trivial to find 
those instabilities which are relevant for the parton system produced 
in ultrarelativistic heavy-ion collisions. 

A system of two interpenetrating beams of nucleons \cite{Iva87,Hen88} or 
partons \cite{Pok88,Mro88c,Pok90a,Pok90b,Pav92} was argued to be unstable with 
respect to the so-called filamentation or Weibel instability \cite{Wei59}. 
However, such a system  appears to be rather unrealistic from the 
experimental point of view. Then, we have argued \cite{Mro93,Mro94b,Mro97b} 
that the filamentation can occur under weaker conditions which are very 
probable for heavy-ion collisions at RHIC and LHC. Instead of the two streams 
of partons, it appears sufficient to assume a strongly anisotropic momentum 
distribution. We systematically review here the whole problem. 

\section{Fluctuation spectrum}

We start with the discussion on how the unstable modes are initiated. 
Specifically, we show that the fluctuations, which act as seeds of the 
filamentation, are {\it large}, much larger than in the equilibrium plasma. 
Since the system of interest is far from the equilibrium, the fluctuations 
are not determined by the chromodielectric permeability tensor discussed in
the previous section. The fluctuation-dissipation theorem does not hold in 
such a case. Thus, we derive the color current correlation function which 
provides the fluctuation spectrum. 

QGP is assumed to be on average locally colorless, homogeneous and 
stationary. Therefore, the distribution functions averaged 
over ensemble are of the form
$$
\langle Q_{ij}(t,{\bf x},{\bf p}) \rangle = 
\delta^{ij} n({\bf p}) \;,\;\;\;\;\;\;
\langle \bar Q_{ij}(t,{\bf x},{\bf p}) \rangle = 
\delta^{ij} \bar n({\bf p}) \;,
$$
$$
\langle G_{ab}(t,{\bf x},{\bf p}) \rangle = 
\delta^{ab} n_g({\bf p}) \;,
$$
which give the zero average color current.

We study the fluctuations of the color current generalizing a well-known
approach to the fluctuating electric current \cite{Akh75}. For a system of 
noninteracting quarks and gluons we have derived (in the classical limit) 
the following expression of the current correlation tensor 
\begin{eqnarray}\label{cur-cor-x}
M^{\mu \nu}_{ab} (t,{\bf x}) &\buildrel \rm def \over =& 
\langle j^{\mu}_a (t_1,{\bf x}_1) j^{\nu}_b (t_2,{\bf x}_2) \rangle 
\nonumber \\
&=& {1 \over 8} \,g^2\; \delta^{ab} 
\int {d^3p \over (2\pi )^3} \; {p^{\mu} p^{\nu} \over E^2} \;
f({\bf p}) \; \delta^{(3)} ({\bf x} -{\bf v} t)  \;,
\end{eqnarray}
where the effective parton distribution function $f({\bf p})$ equals 
$n({\bf p}) + \bar n({\bf p})  +  2N_c n_g({\bf p})$ and 
$(t,{\bf x}) \equiv (t_2-t_1,{\bf x}_2-{\bf x}_1)$. Due to the average 
space-time homogeneity the correlation tensor depends only on the difference 
$(t_2-t_1,{\bf x}_2-{\bf x}_1)$. The physical meaning of the formula 
(\ref{cur-cor-x}) is transparent. The space-time points $(t_1,{\bf x}_1)$ 
and $(t_2,{\bf x}_2)$ are correlated in the system of noninteracting 
particles if the particles fly from $(t_1,{\bf x}_1)$ to $(t_2,{\bf x}_2)$. 
For this reason the delta  $\delta^{(3)} ({\bf x} - {\bf v} t)$ is 
present in the formula (\ref{cur-cor-x}). The momentum integral of the
distribution function simply represents the summation over particles.
The fluctuation spectrum is found as a Fourier transform of the
tensor (\ref{cur-cor-x}) i.e. 
\begin{equation}\label{cur-cor-k}
M^{\mu \nu}_{ab} (\omega ,{\bf k}) = {1 \over 8} \,g^2\; \delta^{ab} 
\int {d^3p \over (2\pi )^3} \; 
{p^{\mu} p^{\nu} \over E^2} \; f({\bf p})  \;
2\pi \delta (\omega -{\bf kv}) \;.
\end{equation}

When the system is in equilibrium the fluctuations are given, according
to the fluctuation-dissipation theorem, by the respective response function.
For $f({\bf p})$ being the classical equilibrium distribution function
one indeed finds the standard fluctuation-dissipation relation \cite{Akh75} 
valid in the $g^2-$order. For example, 
$$
M^{00}_{ab} (\omega ,{\bf k}) = \delta^{ab}  {{\bf k}^2 \over 2\pi} \, 
{T \over \omega} \, {\rm Im} \epsilon_L (\omega ,{\bf k}) \;,
$$
where $T$ is the temperature and $\epsilon_L$ represents the 
longitudinal part of the chromodielectric tensor (\ref{epsL}).

\section{Parton distributions}

We model the parton momentum distribution at the early stage of 
ultrarelativistic heavy-ion collision by two functions:
\begin{equation}\label{f-flat-y}
f({\bf p}) = {1 \over 2Y} 
\Theta(Y - y) \Theta(Y + y) \; h(p_{\bot}) \;
{1 \over p_{\bot} \, {\rm ch}y}\;, 
\end{equation}
and
\begin{equation}\label{f-flat-pl}
f({\bf p}) = {1 \over 2{\cal P}} 
\Theta({\cal P} - p_{\parallel}) 
\Theta({\cal P} + p_{\parallel}) \; 
h(p_{\bot}) \;, 
\end{equation}
where $y$, $p_{\parallel}$ and $p_{\bot}$ denote the parton rapidity, 
the longitudinal and transverse momenta, respectively. The parton momentum 
distribution (\ref{f-flat-y}) corresponds to the rapidity distribution which 
is flat in the interval $(-Y,Y)$. The distribution (\ref{f-flat-pl}) is flat 
for the longitudinal momentum $-{\cal P} < p_{\parallel} < {\cal P}$. We do 
not specify the transverse momentum distribution $h(p_{\bot})$, which is 
assumed to be of the same shape for quarks and gluons, because it is 
sufficient for our considerations to demand that the distributions 
(\ref{f-flat-y}, \ref{f-flat-pl}), are strongly elongated along the 
$z-$axis i.e. $e^Y \gg 1$ and ${\cal P} \gg \langle p_{\bot} \rangle$. 

The QCD-based computations, see e.g. \cite{Gei95,Wan97}, show that the 
rapidity distribution of partons produced at the early stage of heavy-ion 
collisions is essentially gaussian with the width of about one to two units. 
When the distribution (\ref{f-flat-y}) simulates the gaussian one, $Y$ does 
not measure the size of the `plateau' but rather the range over which the 
partons are spread. If one takes the gaussian distribution of the variance 
$\sigma$ and the distribution (\ref{f-flat-y}) of the same variance, then 
$Y = \sqrt{3} \, \sigma $. 

As already mentioned, the parton system described by the distribution 
functions (\ref{f-flat-y}, \ref{f-flat-pl}) is assumed to be homogeneous
and stationary. Applicability of this assumption is very limited because 
there is a correlation between the parton longitudinal momentum and its 
position, i.e. partons with very different momenta will find themselves 
in different regions of space shortly after the collision. However, one 
should remember that we consider the parton system at a very early stage 
of the collision, soon after the Lorentz contracted ultrarelativistic nuclei 
traverse each other. At this stage partons are most copiously produced 
but do not have enough time to escape from each other. Thus, the assumption 
of homogeneity holds for the space-time domain of the longitudinal size, 
say, $2 - 3 \;\; {\rm fm}$ and life time $2 - 3 \;\; {\rm fm}/c$. As shown 
below, this time is long enough for the instability to occur.

\section{Seeds of the filamentation}

Let us now calculate the correlation tensor (\ref{cur-cor-k})
for the distribution functions (\ref{f-flat-y}, \ref{f-flat-pl}). Due to the 
symmetry $f({\bf p}) = f(-{\bf p})$ of these distributions, the tensor 
$M^{\mu \nu}$ is diagonal i.e. $M^{\mu \nu}= 0$ for $\mu \not= \nu$. Since 
the average parton longitudinal momentum is much bigger than the transverse 
one, it obviously follows from Eq.~(\ref{cur-cor-k}) that the largest 
fluctuating current appears along the $z-$axis. Therefore, we discuss the 
$M^{zz}$ component of the correlation tensor. $M^{zz}(\omega, {\bf k})$ 
depends on the ${\bf k}-$vector orientation and there are two generic cases: 
${\bf k} = (k_x,0,0)$ and ${\bf k} = (0,0,k_z)$. The inspection of 
Eq.~(\ref{cur-cor-k}) shows that the fluctuations with ${\bf k} = (k_x,0,0)$ 
are much larger than those with ${\bf k} = (0,0,k_z)$. Thus, let us consider 
$M^{zz}(\omega, k_x)$. Substituting the distributions 
(\ref{f-flat-y}, \ref{f-flat-pl}) into (\ref{cur-cor-k}) one finds after 
azimuthal integration that $M^{zz}_{ab}(\omega, k_x)$ reaches the maximal 
values for $\omega^2 \ll k_x^2$. So, we compute $M^{zz}_{ab}$ at $\omega= 0$. 
Keeping in mind that $e^Y \gg 1$ and ${\cal P} \gg \langle p_{\bot} \rangle$ 
we get the following approximate expressions for the flat $y-$ and 
$p_{\parallel}-$distributions:  
\begin{eqnarray}\label{cor-zzx-0-flat-y}
M^{zz}_{ab}(\omega=0, k_x) = 
{1 \over 8} \,g^2\: \delta^{ab} \; {e^Y \over Y} \;
{\langle \rho \rangle  \over \vert k_x \vert } \;,
\end{eqnarray}
\begin{eqnarray}\label{cor-zzx-0-flat-pl}
M^{zz}_{ab}(\omega=0, k_x) =  
{1 \over 8} \,g^2\: \delta^{ab} \; 
{{\cal P} \over \langle p_{\bot} \rangle } \;
{\langle \rho \rangle  \over \vert k_x \vert } \;,
\end{eqnarray}
where $\langle \rho \rangle$ is the effective parton density given 
for $N_c = 3$ as
\begin{eqnarray*}
\langle \rho \rangle \equiv \int {d^3p \over (2\pi )^3} \; f({\bf p})  
= {1 \over 4\pi^2} \int_0^{\infty} dp_{\bot}p_{\bot} \; h(p_{\bot}) 
= {1\over 3} \; \langle \rho \rangle_{q\bar q} +
{3 \over 4} \; \langle \rho \rangle_g \;,
\end{eqnarray*}
with $\langle \rho \rangle_{q\bar q}$ denoting the average density of 
quarks and antiquarks, and $\langle \rho \rangle_g$ that of gluons.
For the flat $p_{\parallel}-$case we have also used the approximate 
equality
\begin{eqnarray*}
\int_0^{\infty} dp_{\bot} \; h(p_{\bot}) \cong 
{1 \over \langle p_{\bot} \rangle }
\int_0^{\infty} dp_{\bot}p_{\bot} \; h(p_{\bot}) 
\end{eqnarray*}
to get the expression (\ref{cor-zzx-0-flat-pl}). It is instructive to compare 
the results (\ref{cor-zzx-0-flat-y}, \ref{cor-zzx-0-flat-pl}) with the 
analogous one for the equilibrium plasma which is
\begin{eqnarray*}
M^{zz}_{ab}(\omega=0, k_x) = 
{\pi \over 16} \,g^2\; \delta^{ab}\;
{ \langle \rho \rangle \over \vert k_x \vert } \;. 
\end{eqnarray*}
One sees that the current fluctuations in the anisotropic plasma 
are amplified by the {\it large} factor which is $e^Y/Y$ or 
${\cal P} /\langle p_{\bot} \rangle$. With the estimated value of
$Y$ 2.5 for RHIC and 5.0 for LHC \cite{Bir92}, the amplification factor 
$e^Y/Y$ equals 4.9 and 29.7, respectively.

\section{Filamentation mechanism}

Following \cite{Che84} we are going to argue that the fluctuation, 
which contributes to \break $M^{zz}_{ab}(\omega=0,k_x)$, grows in time. The 
form of the fluctuating current is
\begin{eqnarray}\label{flu-cur}
{\bf j}_a(x) = j_a \: \hat {\bf e}_z \: {\rm cos}(k_x x) \;,
\end{eqnarray}
where $\hat {\bf e}_z$ is the unit vector in the $z-$direction.
Thus, there are current filaments of the thickness $\pi /\vert k_x\vert$ 
with the current flowing in the opposite directions in the neighboring 
filaments. For the purpose of a qualitative argumentation presented here 
the chromodynamics is treated as an eightfold electrodynamics. Then, the 
magnetic field generated by the current (\ref{flu-cur}) is given as
\begin{eqnarray*}
{\bf B}_a(x) = {j_a \over k_x} \: \hat {\bf e}_y \: {\rm sin}(k_x x) \;,
\end{eqnarray*}
while the Lorentz force acting on the partons, which fly along the beam, equals
\begin{eqnarray*}
{\bf F}(x) = q_a \: {\bf v} \times {\bf B}_a(x) = 
- q_a \: v_z \: {j_a \over k_x} \: \hat {\bf e}_x \: {\rm sin}(k_x x) \;,
\end{eqnarray*}
where $q_a$ is the color charge. One observes, see Fig.~4, that the force 
distributes the partons in such a way that those, which positively 
contribute to the current in a given filament, are focused to the 
filament center while those, which negatively contribute, are moved 
to the neighboring one. Thus, the initial current grows. 

\section{Dispersion equation}

We analyse here the dispersion equation which for the anisotropic 
plasma is of the form (\ref{disp-eq-det}). The plasma is assumed to be
collisionless i.e. the mean-field interaction dominates the system 
dynamics and $\nu = i0^+$. The assumption is correct if the inverse 
characteristic time of the mean-field phenomena $\tau^{-1}$ is substantially 
larger than the collision frequency $\nu$. Otherwise, the infinitesimally 
small imaginary quantity $i0^+$ from (\ref{epsilon}) should be substituted by 
$i\nu$. Such a substitution however seriously complicates analysis of the 
dispersion equation (\ref{disp-eq-det}). Therefore, we solve the problem 
within the collisionless limit and only {\it a posteriori } argue validity 
of this approximation.

As already mentioned, the solutions $\omega ({\bf k})$ of (\ref{disp-eq-det}) 
are stable when ${\rm Im}\omega < 0$ and unstable when ${\rm Im}\omega > 0$.
It appears difficult to find the solutions of Eq. (\ref{disp-eq-det}) because 
of the complicated structure of the chromodielectric tensor (\ref{epsilon}).
However, the problem simplifies because we are interested in the specific
modes with the wave vector ${\bf k}$ perpendicular and to the chromoelectric 
field ${\bf E}$ parallel to the beam. Thus, we consider the configuration
\begin{equation}\label{configur}
{\bf E} = (0,0,E_z) \;,\;\;\;\; {\bf k} = (k_x,0,0) \;. 
\end{equation}
Then, the dispersion equation (\ref{disp-eq-det}) gets the form
\begin{equation}\label{H-eq}
H(\omega) \equiv k^2_x - \omega^2 \epsilon^{zz}(\omega, k_x) = 0 \;,
\end{equation}
where only one diagonal component of the dielectric tensor enters.

\section{Penrose criterion}

The stability analysis can be performed without solving Eq. 
(\ref{H-eq}) explicitly. Indeed, the so-called Penrose criterion 
\cite{Kra73} states that {\it the dispersion equation $H(\omega )=0$ 
has unstable solutions if} $H(\omega = 0) < 0$. The meaning of this 
statement will be clearer after we will approximately solve the 
dispersion equation in the next section.

Let us compute $H(0)$ which can be written as
\begin{equation}\label{H-zero}
H(0) = k^2_x - \chi^2 \;, 
\end{equation}
with
\begin{equation}\label{chi2}
\chi^2 \equiv - \omega^2_0 
-{g^2 \over 2}\int {d^3p \over (2\pi )^3} \;
{v_z^2 \over v_x} \; {\partial f({\bf p}) \over \partial p_x} \;,
\end{equation}
where the plasma frequency parameter is
\begin{equation}\label{plasma-freq}
\omega^2_0 \equiv - 
{g^2 \over 2} \int {d^3p \over (2\pi )^3} \;
v_z \; {\partial f({\bf p}) \over \partial p_z}
\;. 
\end{equation}
As we shall see below, $\omega_0$ gives the frequency of the stable mode
of the configuration (\ref{configur}) when $k_x \rightarrow 0$.

Substituting the distribution functions (\ref{f-flat-y}, \ref{f-flat-pl}) 
into Eqs.~(\ref{chi2}) and (\ref{plasma-freq}) one finds the analytical but 
rather complicated expression of $H(0)$. In the case of the flat 
$y-$distribution we thus take the limit $e^Y \gg 1$, while for the 
flat $p_{\parallel}-$distribution we assume that 
${\cal P} \gg \langle p_{\bot} \rangle$. Then, we get for the 
flat $y-$distribution
\begin{equation}\label{chi2-y}
\chi^2 \cong - {\alpha_s \over 4 \pi } \; {e^Y \over Y} \; 
\int dp_{\bot} \Bigg( h(p_{\bot}) + 
p_{\bot} {d h(p_{\bot}) \over dp_{\bot}} \Bigg) 
= {\alpha_s \over 4 \pi } \; {e^Y \over Y} \; 
p_{\bot}^{\rm min} h(p_{\bot}^{\rm min}) \;,
\end{equation}
and for the flat $p_{\parallel}-$distribution
\begin{equation}\label{chi2-pl}
\chi^2 \cong - {\alpha_s \over 4 \pi } \; {\cal P} \; 
\int dp_{\bot}  {d h(p_{\bot}) \over dp_{\bot}} 
= {\alpha_s \over 4 \pi } \; {\cal P} \; 
h(p_{\bot}^{\rm min}) \;, 
\end{equation}
where $\alpha_s \equiv g^2 / 4 \pi^2$ is the strong coupling constant and
$p_{\bot}^{\rm min}$ denotes the minimal transverse momentum. The function 
$h(p_{\bot})$ is assumed to decrease faster than $1/p_{\bot}$ when 
$p_{\bot} \rightarrow \infty$. 

As seen, the sign of $H(0)$ given by Eq.~(\ref{H-zero}) is (for sufficiently 
small $k^2_x$) determined by the transverse momentum distribution at the 
minimal momentum. There are unstable modes if 
$p_{\bot}^{\rm min} h(p_{\bot}^{\rm min})>0$ for the flat $y-$distribution 
and if $h(p_{\bot}^{\rm min}) > 0$ for the flat $p_{\parallel}$ case. 
Since the distribution $h(p_{\bot})$ is expected to be a monotonously 
decreasing function of $p_{\bot}$, the instability condition for the flat 
$p_{\parallel}-$distribution seems to be always 
satisfied. The situation with the flat $y-$distribution is less clear. 
So, let us discuss it in more detail. We consider three characteristic 
cases of $h(p_{\bot})$ discussed in the literature.
\begin{enumerate}
\item
The transverse momentum distribution due to a single binary parton--parton 
interaction is proportional to $p_{\bot}^{-6}$ \cite{Esk89} and blows up when 
$p_{\bot} \rightarrow 0$. In such a case 
$p_{\bot}^{\rm min}h(p_{\bot}^{\rm min}) > 0$, there are unstable modes 
and $p_{\bot}^{\min}$ should be treated as a cut-off parameter reflecting 
e.g. the finite size of the system. 

\item
The transverse momentum distribution proportional to 
$(p_{\bot} + m_{\bot})^{-6.4}$ with $m_{\bot} = 2.9$ GeV has been found in 
\cite{Bir92}, where except the binary parton--parton scattering the initial 
and final state radiation has been taken into account. This distribution, 
in contrast to that from 1), gives $p_{\bot}^{\rm min} h(p_{\bot}^{\rm min}) 
= 0$ for $p_{\bot}^{\rm min}=0$ and there is no instability. Although 
one should remember that the finite value of $m_{\bot}$ found in \cite{Bir92} 
is the result of infrared cut-off. Thus, it seems more reasonable to use the 
distribution from 1), where the cut-off explicitly appears.

\item
One treats perturbatively only partons with $p_{\bot} > p_{\bot}^{\rm min}$ 
assuming that those with lower momenta form colorless clusters or strings 
due to a nonperturbative interaction. It should be stressed that the colorless 
objects do not contribute to the dielectric tensor (\ref{epsilon}), which
is found in the {\it linear} response approximation \cite{Mro88c,Mro89}. Thus, 
only the partons with $p_{\bot} > p_{\bot}^{\rm min}$ are of interest for us. 
Consequently  $p_{\bot}^{\rm min} h(p_{\bot}^{\rm min})$ is positive and there 
are unstable modes.  As shown in \cite{Mro94b}, the screening lengths
due to the large parton density are smaller than the confinement scale in 
the vacuum. Therefore, the cut-off parameter $p_{\bot}^{\rm min}$ should
be presumably reduced from 1 - 2 GeV usually used for proton--proton
interactions to, let us say,  0.1 - 0.2 GeV. 

\end{enumerate}

We cannot draw a firm conclusion but we see that the instability condition 
is trivially satisfied for the flat $p_{\parallel}-$distribution and 
is also fulfilled for the flat $y-$distribution under plausible 
assumptions. Let us mention that the difference between the instability
conditions for the flat $y-$ and $p_{\parallel}-$distribution is due
to a very specific property of the $y-$distribution which is limited to the
interval $(-Y,Y)$. The point is that $y \rightarrow \pm \infty$ when 
$p_{\bot} \rightarrow 0$ and consequently, the limits in the rapidity 
suppress the contribution from the small transverse momenta to the dielectric
tensor. For this reason we need for the instability the distribution 
$h(p_{\bot})$ which diverges for $p_{\bot} \rightarrow 0$ in the case of 
the flat $y-$distribution, while the instability condition for the flat 
$p_{\parallel}-$distribution is satisfied when $h(0)$ is finite. If we 
assumed the gaussian rapidity distribution instead of (\ref{f-flat-y}),  
the instability condition would be less stringent. In any case, we assume 
that the Penrose criterion is satisfied  and look for the unstable modes 
solving the dispersion equation (\ref{H-eq}).

\section{Unstable mode}

The dispersion equation (\ref{H-eq}) for a cylindrically symmetric 
system is
\begin{eqnarray}\label{eq-dis-cyl}
k^2_x - \omega^2 + \omega_0^2 &-& {\alpha_s \over 4 \pi^2}
\int_0^{\infty} dp_{\bot}  \int_{-\infty}^{\infty}
{dp_{\parallel} \, p_{\parallel}^2 \over \sqrt{p_{\parallel}^2 + p_{\bot}^2}} 
\nonumber \\
& \times & {\partial f \over \partial p_{\bot}} 
\int_0^{2\pi} 
{d\phi \; {\rm cos}\phi \over a - {\rm cos}\phi + i0^+} = 0 \;, 
\end{eqnarray}
with the plasma frequency $\omega_0$ given by Eq.~(\ref{plasma-freq}) and $a$
denoting
$$
a \equiv {\omega \over k_x} \;
{\sqrt{p_{\parallel}^2 + p_{\bot}^2} \over p_{\bot}}  \;.
$$

We solve Eq.~(\ref{eq-dis-cyl}) in the two limiting cases 
$\vert \omega /k_x \vert \gg 1$ and $\vert k_x /\omega \vert \gg 1$. In the 
first case the azimuthal integral is approximated as 
$$
\int_0^{2\pi} {d\phi \; {\rm cos}\phi \over 
a - {\rm cos}\phi + i0^+} = {\pi \over a^2} + 
{\cal O}(a^{-4}) \;. 
$$
Then, the equation (\ref{eq-dis-cyl}) gets the form
\begin{eqnarray}\label{eq-dis-app}
k^2_x - \omega^2 + \omega^2_0 + \eta^2 {k^2_x \over \omega^2} = 0 \;, 
\end{eqnarray}
where $\eta$, as $\omega_0$, is a constant defined as
$$
\eta^2_0 \equiv -  {\alpha_s\over 4 \pi} \int dp_{\parallel} dp_{\bot}
{ p_{\parallel}^2 p_{\bot}^2 \over (p_{\parallel}^2 + p_{\bot}^2)^{3/2}} 
{\partial f({\bf p}) \over \partial p_{\bot}} \;. 
$$

We have computed $\omega_0$ and $\eta$ for the flat $p_{\parallel}- \;$ and 
$y-$distribution. In the limit $e^Y \gg 1$ and 
${\cal P} \gg \langle p_{\bot} \rangle$, respectively, 
we have found
\begin{equation}\label{omega-y}
\omega_0^2 \cong {\alpha_s \over 8 Y} \int dp_{\bot} h(p_{\bot}) \;, 
\end{equation}
\begin{equation}\label{omega-pl}
\omega_0^2 \cong {\alpha_s \over 2\pi {\cal P}} 
\int dp_{\bot} p_{\bot} h(p_{\bot}) \;, 
\end{equation}
and
\begin{equation}\label{eta-y}
\eta^2 \cong {\alpha_s \over 16 Y} \; 
\int dp_{\bot} \Bigg( {1\over 4} h(p_{\bot}) - p_{\bot} 
{d h(p_{\bot}) \over dp_{\bot}} \Bigg) \;. 
\end{equation}
\begin{equation}\label{eta-pl}
\eta^2 \cong -{\alpha_s \over 4\pi {\cal P}} \; 
{\rm ln}\Bigg({{\cal P} \over \langle p_{\bot} \rangle} \Bigg)
\int dp_{\bot} p_{\bot}^2 {d h(p_{\bot}) \over dp_{\bot}} \;.
\end{equation}

The solutions of Eq. (\ref{eq-dis-app}) are 
\begin{eqnarray}\label{sol-min}
\omega^2_{\pm} = {1 \over 2} \Big( k^2 + \omega^2_0 \pm 
\sqrt{(k^2_x + \omega^2_0)^2 +4 \eta^2 k^2_x} \; \Big) \;. 
\end{eqnarray}
One sees that $\omega^2_+ \ge 0$ and $\omega^2_- \le 0$. Thus, there 
is a pure real mode $\omega_+$, which is stable, and two pure imaginary 
modes $\omega_-$, one of them being unstable. As mentioned previously, 
$\omega_+ = \omega_0 $ when $k_x = 0$.

Let us focus our attention on the unstable mode which can be approximated as 
\begin{displaymath}
\omega_-^2\cong \left\{ 
\begin{array}{ccl}
- {\eta^2 \over \omega_0^2}k^2_x \;, &{\rm for}&  k_x^2 \ll \omega_0^2 \\  
- \eta^2 \;, &{\rm for}& k^2_x \gg \omega_0^2 
\end{array}   \right.
\end{displaymath}
One should keep in mind that Eq. (\ref{sol-min}) holds only for 
$\vert \omega /k_x \vert \gg 1$. We see that $\omega_-$ can satisfy
this condition for $k^2_x \ll \omega_0^2$ if $\eta^2 \gg \omega_0^2$ 
and for $k^2_x \gg \omega_0^2$ if $\eta^2 \ll \omega_0^2$.
To check whether these conditions can be satisfied, we compare 
$\eta^2$ to $\omega_0^2$. Assuming that $h(p_{\bot}) \sim p_{\bot}^{-\beta}$, 
one finds from Eqs.~(\ref{eta-y}, \ref{eta-pl}) 
\begin{equation}\label{eta2-beta}
\eta^2 \cong {1 + 4\beta \over 8} \; \omega_0^2 \;.
\end{equation}
$$
\eta^2 \cong {\beta \over 2}
{\rm ln}\Bigg({{\cal P} \over \langle p_{\bot} \rangle} \Bigg)
 \omega_0^2 \;. 
$$
Since $\beta \cong 6$ \cite{Esk89,Bir92} we get $\eta^2 \geq 3 \omega_0^2$.
Therefore, the solution (\ref{sol-min}) for $k^2_x \ll \omega_0^2$ should be 
correct. 

Let us now  solve the dispersion equation (\ref{eq-dis-cyl}) in the second 
case when $\vert k_x/ \omega \vert \gg 1$. Then, the azimuthal integral from 
Eq.~(\ref{eq-dis-cyl}) is approximated as
$$
\int_0^{2\pi} {d\phi \; {\rm cos}\phi \over 
a - {\rm cos}\phi + i0^+} =  - 2\pi  + {\cal O}(a) \;, 
$$
and we immediately get the dispersion relation
\begin{equation}\label{dis-rel}
\omega^2 =  k^2_x - \chi^2 \;, 
\end{equation}
with $\chi^2$ given by Eq.~(\ref{chi2-y}) or (\ref{chi2-pl}). As previously 
we have assumed that $e^Y \gg 1$ and 
${\cal P} \gg \langle p_{\bot} \rangle$. 
Eq.~(\ref{dis-rel}) provides a real mode for $k^2_x > \chi^2$ and two 
imaginary modes for $k^2_x < \chi^2$. Since the solution (\ref{dis-rel}) 
must satisfy the condition $\vert k_x / \omega \vert \gg 1$, it holds only for 
$k^2_x \gg \vert k^2_x - \chi^2 \vert$.

The dispersion relation of the unstable mode in the whole domain
of wave vectors is schematically shown in Fig.~5, where the solutions 
(\ref{sol-min}) and (\ref{dis-rel}) are combined. Now one sees how the 
Penrose criterion works. When $\chi^2 = 0$ the unstable mode disappears.

\section{Time scales}

The instability studied here can occur in heavy-ion collisions
if the time of instability development is short enough, shorter than 
the characteristic time of evolution of the nonequilibrium state described
by the distribution functions (\ref{f-flat-y}, \ref{f-flat-pl}). 

Let us first estimate the time of instability development which is given
by $1/{\rm Im}\omega$. As seen in Fig.~5, 
$\vert {\rm Im} \omega \vert < \eta$. Thus, we define the minimal
time as $\tau_{\rm min} = 1/\eta$. To find $\tau_{\rm min}$ we estimate the 
plasma frequency. We consider here only the flat $y-$distribution which seems 
to be more reasonable than the flat $p_{\parallel}-$distribution. 
Approximating  $\int dp_{\bot} h(p_{\bot})$ as  
$\int dp_{\bot} p_{\bot} h(p_{\bot})/\langle p_{\bot} \rangle$ 
the plasma frequency (\ref{plasma-freq}) can be written as
\begin{equation}\label{plasma-num}
\omega_0^2 \cong {\alpha_s \pi \over 6 Y r_0^2 A^{2/3}}
(N_q + N_{\bar q} + {9 \over 4} N_g )\;, 
\end{equation}
where $N_c=3$; $N_q$, $N_{\bar q}$ and $N_g$ are the numbers of quarks, 
antiquarks and gluons, respectively, produced in the volume, which has been
estimated in the following way. Since we are interested in the central 
collisions, the volume corresponds to a cylinder of the radius 
$r_0 A^{1/3}$ with $r_0 = 1.1$ fm and $A$ being the mass number of
the colliding nuclei. Using the uncertainty principle argument, 
the length of the cylinder has been taken as $1/\langle p_{\bot} \rangle$, 
which is the formation time of parton with the transverse momentum  
$\langle p_{\bot} \rangle$. 

Neglecting quarks and antiquarks in Eq.~(\ref{plasma-num}) and substituting 
there $N_g = 570$ for the central Au--Au collision at RHIC ($Y = 2.5$) and 
$N_g = 8100$ for the same colliding system at LHC ($Y=5.0$) \cite{Bir92}, 
we get
$$
\omega_0 = 280 \;\; {\rm MeV} \;\;\;{\rm for \;\;RHIC} \;,\;\;\;\;\;\;
\omega_0 = 430 \;\; {\rm MeV} \;\;\;{\rm for \;\;LHC} \;
$$
for $\alpha_s =0.3$ at RHIC and $\alpha_s =0.1$ at LHC. Using 
Eq. (\ref{eta2-beta}) with $\beta = 6$ one finds
$$
\tau_{\rm min} = 0.4 \;\; {\rm fm}/c \;\;\;{\rm for \;\;RHIC} \;,\;\;\;\;\;\;
\tau_{\rm min} = 0.3 \;\; {\rm fm}/c \;\;\;{\rm for \;\;LHC} \;. 
$$

The plasma has been assumed collisionless in our analysis. Such an assumption
is usually correct for weakly interacting systems because the damping 
rates of the collective modes due to collisions are of the higher order in 
$\alpha_s$ than the frequencies of these modes, see e.g. \cite{Akh75}. 
However, it has been argued recently \cite{Gyu93} that the color collective 
modes are overdamped due to the unscreened chromomagnetic interaction. 
However it is unclear whether these arguments concern the unstable mode 
discussed here. The point is that the paper \cite{Gyu93} deals with the 
neutralization of color charges which generate the longitudinal chromoelectric 
field while the unstable mode which we have found is transversal and 
consequently is generated by the color currents not charges. Let us refer 
here to the electron-ion plasma, where the charge neutralization is a very 
fast process while currents can exist in the system for a much longer time 
\cite{Kra73}. In any case, the above estimates of the instability development 
should be treated as lower limits.

Let us now discuss the characteristic time of evolution of the nonequilibrium 
state described by the distribution functions (\ref{f-flat-y}, \ref{f-flat-pl}). 
Except the possible unstable collective modes, there are two other important 
processes responsible for the temporal evolution of the initially produced 
many-parton system: free streaming \cite{Hwa86,Bir93,Esk93} and parton--parton 
scattering. The two processes lead to the isotropic momentum distribution of 
partons in a given cell. The estimated time to achieve local isotropy 
due to the free streaming is about 0.7 fm/c at RHIC \cite{Esk93}. 
The estimates of the equilibration time due to the parton scatterings 
are similar \cite{Gei92,Shu87}. As seen the three time scales of interest are 
close to each other.  Therefore, the color unstable modes can play a role 
in the dynamics of many-parton system produced at the early stage of 
heavy-ion collision, but presumably the pattern of instability cannot 
fully develop.

\section{Detecting the filamentation}

One asks whether the color instabilities are detectable in 
ultrarelativistic heavy-ion collisions. The answer seems to be positive 
because the occurrence of the filamentation breaks the azimuthal symmetry 
of the system and hopefully will be visible in the final state. The azimuthal 
orientation of the wave vector will change from one collision to another while 
the instability growth will lead to the energy transport along this vector 
(the Poynting vector points in this direction). Consequently, one expects 
significant variation of the transverse energy as a function of the azimuthal
angle. This expectation is qualitatively different than that based on the 
parton cascade simulations \cite{Gei95}, where the fluctuations are strongly 
damped due to the large number of uncorrelated partons. Due to the collective 
character of the filamentation instability the azimuthal symmetry will be 
presumably broken by a flow of large number of particles with relatively 
small transverse momenta. The jets produced in hard parton-parton 
interactions also break the azimuthal symmetry. However, the symmetry is 
broken in this case due to a few particles with large transverse momentum.
The problem obviously needs further studies but the event-by-event analysis 
of the nuclear collision seems to give a chance to observe the color 
instabilities in the experiments planed at RHIC and LHC.

\newpage
\vspace{1cm}
\begin{center}
{\bf Figure Captions}
\end{center}
\vspace{0.3cm}
\noindent
{\bf Fig. 1.} 
The contour along the time axis for an evaluation of the
operator expectation values.

\vspace{0.5cm}

\noindent
{\bf Fig. 2.} 
The lowest-order diagram of the self-energy.

\vspace{0.5cm}

\noindent
{\bf Fig. 3.} 
The second-order diagrams of the self-energy.

\vspace{0.5cm}

\noindent
{\bf Fig. 4.} The mechanism of filamentation. The phenomenon is, for
simplicity, considered in terms of the electrodynamics. The fluctuating 
current generates the magnetic field acting on the positively charged 
particles which in turn contribute to the current (see text). 
$\otimes$ and $\odot$ denote the parallel and, respectively, antiparallel 
orientation of the magnetic field with respect to the $y-$axis.

\vspace{0.5cm} 

\noindent
{\bf Fig. 5.} The schematic view of the dispersion relation of 
the filamentation mode.


\newpage

\begin{thebibliography}{99}

\bibitem{QM96} Proc. of Int. Conf. on Ultrarelativistic Nucleus-Nucleus
Collisions {\it Quark Matter'96}, Heidelberg, 1996, edited by 
P. Braun-Munzinger, H.J. Specht, R. Stock and H. St\" ocker (North-Holland, 
Amsterdam, 1996) Nucl. Phys. {\bf A610} (1996).

\bibitem{Akh75} A.I. Akhiezer, I.A. Akhiezer, R.V. Polovin, A.G. Sitenko
and K.N. Stepanov, {\it Plasma Electrodynamics} (Pergamon, New York, 1975).

\bibitem{Bez72} B. Bezzerides and D.F. Dubois, Ann. Phys. {\bf 70} (1972) 10.

\bibitem{Bia88a} A. Bia\l as and W. Czy\. z, Ann. Phys. {\bf 187} (1988) 97.

\bibitem{Bia88b} A. Bia\l as, W. Czy\. z, A. Dyrek and 
W. Florkowski, Nucl. Phys. {\bf B296} (1988) 611.

\bibitem{Bir92} T.S. Bir\' o, B. M\" uller and X.-N. Wang, Phys. Lett. 
{\bf B283} (1992) 171.

\bibitem{Bir93} T.S. Bir\' o {\it et al.}, Phys. Rev. {\bf C48} (1993) 1275.

\bibitem{Bjo64} J. D. Bjorken and S. D. Drell, {\it Relativistic Quantum 
Fields} (McGraw-Hill, San Francisco, 1964).

\bibitem{Bla94} J.-P. Blaizot and E. Iancu, Nucl. Phys. {\bf B417} (1994) 608.

\bibitem{Boy96} D. Boyanowsky, I.D. Lawrie and D.-S. Lee, 
Phys. Rev. {\bf D54} (1996) 4013.

\bibitem{Bra90} E. Braaten and R. D. Pisarski, Nucl. Phys. 
{\bf B337}, 569 (1990).

\bibitem{Cal88} E. Calzetta and B. L. Hu, Phys. Rev. {\bf D37} (1988) 2878.

\bibitem{Che84} F.F. Chen, {\it Introduction to Plasma Physics
and Controlled Fusion} (Plenum Press, New York, 1984).

\bibitem{Dan84} P. Danielewicz, Ann. Phys. (N.Y.) {\bf 152} (1984) 239.

\bibitem{Elz86a} H.-Th. Elze, M. Gyulassy and D. Vasak, Nucl. Phys.
{\bf B276} (1986) 706.

\bibitem{Elz86b} H.-Th. Elze, M. Gyulassy and D. Vasak, Phys. Lett.
{\bf B177} (1986) 402.

\bibitem{Elz87} H.-Th. Elze, Z. Phys. {\bf C38} (1987) 211.

\bibitem{Elz89} H.-Th. Elze and U. Heinz, Phys. Rep. {\bf 183} (1989) 81.

\bibitem{Esk89} K.J. Eskola, K. Kajantie and J. Lindfors, Nucl. Phys. 
{\bf B323} (1989) 37.

\bibitem{Esk93} K.J. Eskola and X.-N. Wang, Phys. Rev. {\bf C47} (1993) 2329.

\bibitem{Gei92} K. Geiger, Phys. Rev. {\bf D46} (1992) 4986. 

\bibitem{Gei95} K. Geiger, Phys. Rep. {\bf 258} (1995) 237.

\bibitem{Gei96} K. Geiger, Phys. Rev. {\bf D54} (1996) 949.

\bibitem{Gro80} S.R. deGroot, W.A. van Leeuwen and Ch. G. van Weert,
{\it Relativistic Kinetic Theory} (North-Holland, Amsterdam, 1980).

\bibitem{Gyu93} M. Gyulassy and A.V. Selikhov, Phys. Lett. 
{\bf B316} (1993) 373.

\bibitem{Han87} T.H. Hansson and I. Zahed, Nucl. Phys., {\bf B292} (1987) 725.

\bibitem{Hei83} U. Heinz, Phys. Rev. Lett., {\bf 51} (1983) 351.

\bibitem{Hei84} U. Heinz, Nucl. Phys. {\bf A418} (1984) 603c.

\bibitem{Hei85a} U. Heinz, Ann. Phys. {\bf 161} (1985) 48.

\bibitem{Hei85b} U. Heinz and P.J. Siemens, Phys. Lett. {\bf B158} (1985) 11.

\bibitem{Hei86} U. Heinz, Ann. Phys. {\bf 168} (1986) 148.

\bibitem{Hei87} U. Heinz, K. Kajantie and T. Toimela, Ann. Phys. 
{\bf 176} (1987) 218.

\bibitem{Hen88} P. Henning and B.L. Friman, Nucl. Phys. {\bf A490} (1988) 689.

\bibitem{Hen90} P. Henning, Nucl. Phys. {\bf B337} (1990) 547.

\bibitem{Hen95} P. Henning, Phys. Rep. {\bf 253} (1995) 235.

\bibitem{Hwa86} R. Hwa and K. Kajantie, Phys. Rev. Lett. {\bf 56} (1986) 696.

\bibitem{Iva87} Yu.B. Ivanov, Nucl. Phys. {\bf A474} (1987) 693.

\bibitem{Kad62} L.P. Kadanoff and G. Baym, 
{\it Quantum Statistical Mechanics} (Benjamin, New York, 1962).

\bibitem{Kal84} O.K. Kalashnikov, Fortschritte Phys. {\bf 32} (1984) 525.

\bibitem{Kel64} L.V. Keldysh, Zh. Eksp. Teor. Fiz. {\bf 47} (1964) 1515.

\bibitem{Kel94} P.F. Kelly, Q. Liu, C. Lucchesi and C. Manuel, 
Phys. Rev. {\bf D50} (1994) 4209.

\bibitem{Kle97} S.P. Klevansky, A. Ogura and J. H\" ufner, 
Ann. Phys. {\bf 261} (1997) 37.

\bibitem{Kra73}  N.A. Krall and A.W. Trivelpiece, {\it Principles of Plasma
Physics} (McGraw-Hill, New York, 1973). 

\bibitem{Lan60} L.D. Landau and E.M. Lifshitz, {\it Electrodynamics of
Continuous Media} (Pergamon, New York, 1960).

\bibitem{Li83} S.-P. Li and L. McLerran, Nucl. Phys. {\bf B214} (1983) 417.

\bibitem{Lif81} E.M. Lifshitz and Pitaevskii, {\it Physical Kinetics}
(Pergamon, New York, 1981).

\bibitem{Mro87a} St. Mr\' owczy\' nski, Phys. Lett. {\bf B188} (1987) 127.

\bibitem{Mro88c} St. Mr\' owczy\' nski, Phys. Lett. {\bf B214} (1988) 587.

\bibitem{Mro89} St. Mr\' owczy\' nski, Phys. Rev. {\bf D39} 1940 (1989). 

\bibitem{Mro90} St. Mr\' owczy\' nski and P. Danielewicz, 
Nucl. Phys. {\bf B342} (1990) 345.

\bibitem{Mro93} St. Mr\' owczy\' nski, Phys. Lett. {\bf B314} (1993) 118.

\bibitem{Mro94a} St. Mr\' owczy\' nski and U. Heinz, 
Ann. Phys. {\bf 229} (1994) 1.

\bibitem{Mro94b} St. Mr\' owczy\' nski, Phys. Rev. {\bf C49} (1994) 2191. 

\bibitem{Mro97a} St. Mr\' owczy\' nski, Phys. Rev. {\bf D56} (1997) 2265.

\bibitem{Mro97b} St. Mr\' owczy\' nski, Phys. Lett. {\bf B393} (1997) 26. 

\bibitem{Nor28} L.W. Nordheim, Proc. Roy. Soc. (London) {\bf A119} (1928) 689. 

\bibitem{Pok88} Yu. E. Pokrovskii and A.V. Selikhov, Pis'ma Zh. Eksp.
Teor. Fiz. {\bf 47} (1988) 11.

\bibitem{Pok90a} Yu. E. Pokrovskii and A.V. Selikhov, Yad. Fiz. 
{\bf 52} (1990) 229.

\bibitem{Pok90b} Yu. E. Pokrovskii and A.V. Selikhov, Yad. Fiz. 
{\bf 52} (1990) 605.

\bibitem{Pav92} O.P. Pavlenko, Yad. Fiz. {\bf 55} (1992) 2239.

\bibitem{Reh98} P. Rehberg, hep-ph/9803239, to appear in Phys. Rev.
{\bf C}.

\bibitem{Sel91} A. V. Selikhov, Phys. Lett. {\bf B268}, 263 (1991).

\bibitem{Sel94} A. V.  Selikhov and M. Gyulassy, 
Phys. Rev. {\bf C49}, 1726 (1994).

\bibitem{Sil61} V.P. Silin and A.A. Ruhadze, {\it Electrodynamics of
Plasma and Plasma-like Media} (Gosatomizdat, Moscow, 1961) (in Russian).

\bibitem{Sch61} J. Schwinger, J. Math. Phys. {\bf 2} (1961) 407.

\bibitem{Shu87} E. Shuryak, Phys. Rev. Lett. {\bf 68} (1992) 3270. 

\bibitem{Ueh33} E.A. Uehling and G.E. Uhlenbeck, Phys. Rev.
{\bf 43} (1933) 552.

\bibitem{Wan97} X.-N. Wang, Phys. Rep. {\bf 280} (1997) 287.

\bibitem{Wei59} E.S. Weibel, Phys. Rev. Lett. {\bf 2} (1959) 83.

\bibitem{Wel82} H.A. Weldon, Phys. Rev. {\bf D26} (1982) 1394.

\bibitem{Wel83} H.A. Weldon, Phys. Rev. {\bf D28} (1983) 2007.

\bibitem{Win84} J. Winter, J. Phys. (Paris) {\bf 45}, C6 (1984) 53.

\bibitem{Ynd83} F.J. Yndurain, {\it Quantum Chromodynamics } 
(Springer, New York, 1983).

\end{thebibliography}
\end{document}